\documentclass[prd,preprintnumbers,twocolumn,eqsecnum,floatfix,letterpaper,superscriptaddress,nofootinbib]{revtex4}
\usepackage{color}
\usepackage{amsmath,amssymb,graphicx}
\usepackage{bm}
\usepackage{times}
\usepackage{microtype}
\usepackage{booktabs}
\usepackage{subfigure}
\usepackage[varg]{txfonts}
\allowdisplaybreaks
\newcommand{\red}[1]{\textcolor{black}{#1}}

\begin{document}

\newcommand{\LNh}{\hat{\mathbf{L}}_\text{N}}
\newcommand{\LN}{\mathbf{L}_\text{N}}
\newcommand{\bS}{\mathbf{S}}
\newcommand{\bJ}{\mathbf{J}}
\newcommand{\e}{\mathrm{e}}
\newcommand{\rmi}{\mathrm{i}}
\newcommand{\flow}{f_0}
\newcommand{\fcut}{f_\mathrm{cut}}

\newcommand{\bchi}{\bm{\chi}}
\newcommand{\blambda}{\bm{\lambda}}
\newcommand{\bLambda}{\bm{\Lambda}}
\newcommand{\bchia}{\bm{\chi}_\text{a}}
\newcommand{\bchis}{\bm{\chi}_\text{s}}
\newcommand{\chis}{\chi_\text{s}}
\newcommand{\chia}{\chi_\text{a}}
\newcommand{\chiadL}{\bchia \cdot \LNh}
\newcommand{\chisdL}{\bchis \cdot \LNh}
\newcommand{\chisSqr}{\bchis^2}
\newcommand{\chiaSqr}{\bchia^2}
\newcommand{\chisDchia}{\bchis \cdot \bchia}
\newcommand{\cA}{\mathcal{A}}
\newcommand{\cB}{\mathcal{B}}
\newcommand{\cC}{\mathcal{C}}
\newcommand{\cP}{\mathcal{P}}
\newcommand{\Mc}{M_\mathrm{c}}
\newcommand{\thetaz}{\theta_{0}}
\newcommand{\thetat}{\theta_{3}}
\newcommand{\thetats}{\theta_\mathrm{3S}}
\newcommand{\psiL}{\psi^\text{L}}
\newcommand{\dphi}{\partial \psi}
\newcommand{\dtheta}{\partial \theta}
\newcommand{\dphiL}{\partial \psi^\text{L}}
\newcommand{\derb}{\partial_b}
\newcommand{\dera}{\partial_a}
\newcommand{\df}{{\mathrm{d}f}}
\newcommand{\match}{\mathcal{M}}
\newcommand{\bOmega}{\mathbf{\Omega}}
\newcommand{\btheta}{\bm{\theta}}
\newcommand{\SBank}{\textsc{SBank}}
\newcommand{\LALSuite}{\textsc{LALSuite}}
\newcommand{\FFe}{\mathrm{FF_{eff}}}
\newcommand{\FF}{\mathrm{FF}}
\newcommand{\bg}{\mathbf{g}}

\newcommand{\LIGO}{LIGO Laboratory, California Institute of Technology,
Pasadena, CA 91125, USA}
\newcommand{\CIT}{Theoretical Astrophysics, California Institute of Technology,
Pasadena, CA 91125, USA}
\newcommand{\ICTS}{International Centre for Theoretical Sciences, Tata Institute of Fundamental Research, Bangalore 560012, India}

\title{An effectual template bank for the detection of gravitational waves\\ from inspiralling compact binaries with generic spins}
\author{P.~Ajith}
\affiliation{\ICTS}
\affiliation{\LIGO}
\affiliation{\CIT}
\author{N.~Fotopoulos}
\affiliation{\LIGO}
\author{S.~Privitera}
\affiliation{\LIGO}
\author{A.~Neunzert}
\affiliation{Oregon State University, Corvallis, OR 97331, USA}
\author{N.~Mazumder}
\affiliation{Indian Institute of Science Education and Research, Thiruvananthapuram 695016, India}
\author{A.~J.~Weinstein}
\affiliation{\LIGO}

\begin{abstract}
We report the construction of a three-dimensional template bank for the search for gravitational waves from inspiralling binaries consisting of spinning compact objects. The parameter space consists of two dimensions describing the mass parameters and one ``reduced-spin'' parameter, which describes the secular (non-precessing) spin effects in the waveform. The template placement is based on an efficient stochastic algorithm and makes use of the semi-analytical computation of a metric in the parameter space. We demonstrate that for ``low-mass''  ($m_1 + m_2 \lesssim 12\,M_\odot$) binaries, this template bank achieves effective fitting factors \red{$\sim0.92$--$0.99$} towards signals from \emph{generic spinning} binaries in the advanced detector era over the entire parameter space of interest (including binary neutron stars, binary black holes, and black hole-neutron star binaries). This provides a powerful and viable method for searching for gravitational waves from generic spinning low-mass compact binaries. Under the assumption that spin magnitudes of black-holes [neutron-stars] are uniformly distributed between 0--0.98 [0 -- 0.4] and spin angles are isotropically distributed, the expected improvement in the average detection volume (at a fixed signal-to-noise-ratio threshold) of a search using this reduced-spin bank is \red{$\sim20-52\%$}, as compared to a search using a non-spinning bank.
\end{abstract}

\pacs{04.25.D, 04.30.-w, 04.25.dg, 29.85.Fj}
\preprint{LIGO-P1200106-v3}
\maketitle

\section{Introduction}
The quest for the first direct detection of gravitational waves (GWs) is entering a new era. The first generation interferometric GW detectors (LIGO, Virgo, GEO\,600 and TAMA) have been decommissioned and are being upgraded to their advanced configurations. Although a detection of GWs is yet to be made, the non-detection is consistent with our expectation of the event rates of the astrophysical phenomena producing GWs of detectable strength (see, e.g.~\cite{Abadie:2010cf}).  The second-generation detectors are expected to be operational in a few years and to reach their design sensitivities within this decade. With roughly an order of magnitude improved strain sensitivity as compared to their first generation counterparts (and hence three orders of magnitude increase in the volume of the local universe accessible to GW observations), the second-generation detectors are expected to make the first detections, opening up a new observational window to the Universe (see, e.g.,~\cite{Aasi:2013wya}). 

Among the most promising sources for the first detection of GWs are the coalescence of astrophysical compact binaries consisting of neutron stars and/or black holes. The rationale for this expectation is at least threefold: 1) Compact binaries are \emph{efficient} sources of GWs. During the coalescence process about $1$--$15$\% of the mass-energy of the binary will be radiated as GWs, which means that such sources can be observed up to very large distances. 2) Radio observations of binary pulsars provide strong observational evidence supporting the existence of at least one class of such sources, i.e., binaries of two neutron stars. 3) These are remarkably ``clean'' sources---the expected GW signals can be accurately modeled and easily parametrized in terms of the component masses and spins. This last point means that the data analysis can benefit from powerful detection techniques such as the \emph{matched filtering}, which is the optimal filter to detect known signals buried in noisy data.

Matched filtering involves cross correlating the data with theoretical templates of the expected signals. Theoretical signal templates can be calculated by employing perturbative or numerical techniques to solve the Einstein's equations of General Relativity~\footnote{Proper calculation of the GW signals from the last stages (merger) of the coalescence of binaries involving neutron stars also requires considering the effect of the nuclear matter, in addition to General Relativity.}. During the early stages of the coalescence, called the \emph{adiabatic inspiral}, the dynamics of the binary as well as the expected gravitational waveforms can be calculated using the post-Newtonian (PN) approximation to General Relativity~\cite{Blanchet:LivRev}. For the case of ``low-mass'' binaries (total mass $\lesssim 12\,M_\odot$), the signal in the frequency band of the ground-based interferometric detectors is dominated by the adiabatic inspiral. Hence it is sufficient to employ waveform templates modelling only the inspiral part of the coalescence, which is well described by the PN theory. On the other hand, for binaries with total mass $\gtrsim 12\,M_\odot$ the ``post-inspiral'' stages (such as the merger and ringdown) also contribute to the signal observed by the ground-based interferometers~\cite{Buonanno:2009zt,Ajith:2007xh}. Thus, the templates have to model the complete inspiral, merger and ringdown. This requires inputs from the numerical-relativity simulations, apart from perturbative General Relativity.

Although the expected waveforms can be computed as a function of the source parameters (such as the component masses and spins), the parameters of the signal that is buried in the data are generally not known. Thus, the data has to be cross-correlated with a ``bank'' of templates corresponding to a discrete set of parameters. The discretization of the parameter space is governed by two requirements: 1) the templates have to be placed with sufficient ``resolution'' in the parameter space such that the loss of signal-to-noise ratio (SNR) due to the mismatch of the closest template to a signal with arbitrary values of the parameters is minimal. 2) the computational cost of the search using the full template bank should be manageable. A geometrical formalism~\cite{Owen:1995tm,Owen:1998dk} has been developed to lay down templates in the parameter space corresponding to a given value of acceptable loss of SNR (or \emph{mismatch}). This is based on defining a metric in the parameter space such that mismatch between two neighboring templates give the proper distance between them~\cite{Owen:1995tm,Owen:1998dk}.

While a closed-form expression of the template-space metric can be computed for the case of binaries with negligible spins (or with spins aligned/anti-aligned with the orbital angular momentum), this is not possible for binaries with generic spins. This is due to the fact that, if the spins are misaligned with the orbital angular momentum of the binary, the spin-orbit and spin-spin interactions will cause the spins (and hence the orbit also) to precess. The resulting dynamics as well as the gravitational waveforms are rather complex, and the modelling requires solving a set of coupled ordinary differential equations. The template placement is further complicated by the large dimensionality of the parameter space (two mass parameters, five spin parameters, and two angles describing the orientation of the binary, in general).

Almost all searches for GWs from coalescing compact binaries using the data of first generation interferometers have used templates that neglect the spins of the compact objects~\cite{Colaboration:2011nz,Abadie:2010yb,Abadie:2011kd,Abbott:2009tt,Abbott:2009qj,Abbott:2007ai}. This was primarily motivated by the expectation that, for the case of first generation detectors, the non-spinning templates are sufficient for the detection of spinning binaries over a significant fraction of the parameter space (loss of detection efficiency is acceptable). Another reason is the lack of a search strategy with improved efficiency (for a given false alarm probability) as compared to the non-spinning search and that is computationally viable~\cite{VanDenBroeck:2009gd}.

The low-frequency sensitivity of the second-generation detectors is expected to be significantly better than that of the first generation detectors. For example, the Advanced LIGO detectors are expected to be sensitive to frequencies above $10$--$20\,$Hz, while the Initial LIGO detectors were only sensitive to frequencies above $40\,$Hz. Hence the second-generation detectors will be able to observe the inspiral from much larger orbital separations, and the observed GW signal will be significantly longer. As a result, neglecting the spin effects can cause a much larger dephasing of the template with the signal, and hence considerable loss of SNR~\cite{Ajith:2011ec,Brown:2012gs}. Proper consideration of the spin effects is essential in the advanced detector era.

The fact that several different spin configurations are nearly degenerate has been recognized for quite some time. Different ideas for the construction of template banks for spinning binaries have been proposed in the past. All these proposals sought to reduce the effective dimensionality of the parameter space by making use of the near-degeneracies \cite{PhysRevD.54.2438,PhysRevD.67.042003,BCV2,Pan:2003qt,Buonanno:2004yd,FaziThesis,Harry:2010fr,Harry:2011qh,Boyle:2009dg,Aylott:2009ya}; see Sec. I of ~\cite{Ajith:2011ec} for a brief summary. However, a computationally viable spinning search that is more efficient than a non-spinning search was demonstrated for the first time only recently~\cite{Privitera:2013xza}, which studied the efficiency (at a fixed false alarm rate) of a search using non-precessing-spin templates towards binaries with non-precessing spins. 
Recently, it has been observed that non-precessing templates are also effectual in detecting binaries with \emph{generic} spins if the mass ratio is moderate ($m_2/m_1 \lesssim 10$)~\cite{Ajith:2011ec}. Furthermore, it was demonstrated that these spin effects can be described by a single \emph{reduced-spin} parameter in an approximate fashion. This opened up the possibility of performing searches for generic spinning binaries using a template family described by just three parameters (two parameters describing the masses and one describing the spins).

In this paper, we extend the previous work presented in Ref.~\cite{Ajith:2011ec}, which proposed a frequency-domain, spinning PN template family described by three parameters. Here we construct a computationally efficient three-dimensional template bank for laying down templates in the parameter space. The template bank is based on an efficient stochastic template-placement algorithm and makes use of a fast, semi-analytical computation of the template-space metric for the waveform family mentioned above. The template-placement algorithm and the computer code that is used for the construction of the template bank is quite generic and can be used for constructing template banks for other waveform families as well.

Finally, we note that other approaches are also being explored towards developing efficient and feasible searches for GWs from spinning compact binaries. Other ideas include allowing the templates to take nonphysical values for the component masses, thus mimicking the effect of spins (see e.g.,~\cite{Aylott:2009ya}), simplifying the waveforms and the metric by neglecting the spin effects of the smaller body~\cite{BCV2,Pan:2003qt,Buonanno:2004yd,FaziThesis,Harry:2010fr,Harry:2011qh}, reducing the effective dimensionality of the parameter space by numerically identifying the principal components~\cite{Brown:2012qf,Harry:2013tca}, etc. The number of templates in a bank can be further reduced by finding a near-orthonormal basis of the template space by employing either the reduced-basis approach~\cite{Herrmann:2012if} or by employing singular-value decomposition of the template bank~\cite{PhysRevD.82.044025}.

The rest of the paper is organized as follows: Sec~\ref{sec:Summary} presents a compact summary of the paper. Sec.~\ref{sec:Waveforms} describes the reduced-spin PN waveform family describing the GWs from binaries with non-precessing spins. Computation of the template-space metric for this waveform family is described in Sec.~\ref{sec:metric}, where we also compare the analytical computation of the metric with exact numerical computations. Sec.~\ref{sec:TemplateBank} discusses the construction of the template bank using a stochastic placement algorithm and the template-space metric. The effectualness of the template bank in detecting generic spinning binaries is demonstrated in Sec.~\ref{sec:Application}, while Sec.~\ref{sec:Conclusions} contain some concluding remarks. We use geometrical units throughout the paper: $G=c=1$.

\subsection{Summary of the paper}
\label{sec:Summary}

A quick summary of this paper is as follows: We present a three-dimensional, stochastic template bank employing templates describing GW signals from compact binaries with non-precessing spins. The parameter space consists of two dimensions describing the mass parameters and one dimension describing a ``reduced-spin'' parameter that describes secular effects of spin. This reduced-spin template bank achieves \emph{effective fitting factor} [see~Eq.(\ref{eq:FF_eff}) for definition] \red{$\sim0.98$} towards binaries with non-precessing spins. Under the assumption that spin magnitudes of black-holes [neutron-stars] are uniformly distributed between 0--0.98 [0 -- 0.4] and spin angles are isotropically distributed, the effective fitting factor of the reduced-spin template bank is \red{$0.92$--$0.99$} towards \emph{generic} spinning binaries with total mass $\lesssim 12\,M_\odot$ (see Fig.~\ref{fig:FFeSpinTaylorT5BankSim}), while the corresponding fitting factor of a {non-spinning} template bank is \red{$0.83$--$0.88$}. A search using the reduced-spin template bank is expected to bring about \red{$20$--$58\%$} increase in the detection volume at a fixed SNR threshold compared to a search using only non-spinning templates (see Fig.~\ref{fig:VolIncr}).

The high effectualness of the reduced-spin template bank (which does not seek to model the modulational effects of precession) is due to the fact that, for the case of comparable-mass binaries ($m_1 \sim m_2$) the total angular momentum of the binary is dominated by the orbital angular momentum, and hence the modulational effects of spin precession on the orbit, and hence on the observed signal, is small. This effect is further enhanced by the intrinsic selection bias towards binaries that are nearly ``face-on'' with the detector (where the modulational effects of precession are weak while the signal is strong) as opposed to binaries that are nearly ``edge-on'' (where the modulational effects are strong while the signal is weak).

The template placement is based on an efficient stochastic algorithm, and makes use of the semi-analytical computation of a metric in the parameter space, which significantly reduces the computational cost. For the range of parameters that we have considered (see Table~\ref{tab:bank_params}) the reduced-spin template bank results in a factor of $\sim 7.5$ increase in the number of templates (as compared to a corresponding non-spinning bank). This number is indicative of the expected increase in the computational cost of the spinning search. We also emphasize that the template placement algorithm and the computer code (called \SBank{}) can be applied to arbitrary template waveforms with arbitrary dimensions and is available in the \LALSuite{} GW data analysis software~\cite{LAL}. 

\section{Reduced-spin waveform templates for inspiralling compact binaries}
\label{sec:Waveforms}

During the early stages of the coalescence of the compact binaries, there is a clean separation of time scales. The orbital time scale (the orbital period) is much shorter than the precession time scale (the time scale over which the spins/orbit precess around the total angular momentum axis), which is much shorter than the inspiral time scale (the time scale over which the orbital separation decreases). i.e.,
\begin{equation}
t_\mathrm{orbit} \ll t_\mathrm{precession} \ll t_\mathrm{inspiral}.
\end{equation}
This clean separation of time scales considerably simplifies the equations to be solved for computing the expected GW signals. The gravitational waveforms can be computed by solving the following set of coupled ordinary differential equations~\cite{PhysRevD.49.6274}:
\begin{eqnarray}
\frac{dE(v)}{dt}   & = & -\frac{\mathcal{F}(v)}{m}, ~~~ \mathrm{(energy~balance)} \nonumber \\
\frac{d \bS_i}{dt} & = & \bOmega_i \times \bS_i \,,~ i = 1,2,  ~~~ \mathrm{(spin~precession)} \nonumber \\
\frac{d \LNh}{dt}  & = & \frac{-1}{||\mathbf{L}||} \, \frac{d}{dt} (\bS_1 + \bS_2). ~~~ \mathrm{(orbital~precession)}
\label{eq:PNEvlEqns}
\end{eqnarray}
The first equation above is the energy-balance argument, which relates the PN expansions of the \emph{specific} binding energy $E(v)$ of the orbit with the GW luminosity ${\mathcal{F}(v)}$.  The PN expansion parameter $v$ is related to the orbital frequency $\omega$ by $v\equiv (m\omega)^{1/3}$, where $m \equiv m_1 + m_2$ is the total mass of the binary. The second equation represents the precession of the spin vectors $\bS_1$ and $\bS_2$, where the magnitudes of the vectors $\bOmega_1$ and $\bOmega_2$ are the spin precession frequencies. In the leading order, $\bOmega_1$ and $\bOmega_2$ are parallel to the Newtonian orbital angular momentum vector $\LN$. The third equation represents the precession of the orbital plane (described by the unit vector $\LNh$ along the Newtonian orbital angular momentum), which is derived from the conservation of the total angular momentum over the precession time scale.  $||\mathbf{L}||$ is the magnitude of the orbital angular momentum $\mathbf{L}$. Due to the precession of the orbital plane, and since the GWs are predominately beamed along the direction of the orbital angular momentum, the GW signal observed by a fixed detector will contain complicated amplitude and phase modulations.

The essence of the reduced-spin templates proposed by~\cite{Ajith:2011ec} is the following: In the case of binaries with moderate mass ratios ($m_2 \sim m_1$), the total angular momentum of the binary is dominated by the orbital angular momentum. Hence the amount of orbital precession required to conserve the total angular momentum (by ``compensating'' for the spin precession) is rather small. As a result, the modulational effects of precession on the waveform are small; the spin effects are nearly secular. Consequently, the phase evolution is very similar to that of a \emph{non-precessing} binary  (spins aligned/anti-aligned to the orbital angular momentum, so that $d\bS_i/dt = 0$ in Eq.~\ref{eq:PNEvlEqns}) with a different value of the spins (see, also~\cite{PhysRevD.89.044021,PhysRevD.88.024040,PhysRevD.86.104063}). This approximate mapping between precessing and non-precessing spins suggest that we will be able to detect some of the precessing binaries employing just non-precessing templates. This approximation (that the spins are non-precessing) has two advantages: 1) This enables us to compute an explicit, closed-form expression of the Fourier transform of the template using the \emph{stationary-phase approximation}~\cite{Poisson:1995ef,Arun:2009}. 2) This enables us to describe the spin effects using a single \emph{reduced-spin} parameter, in an approximate way~\cite{Ajith:2011ec}.

The ``reduced-spin'' waveform templates are defined in the frequency domain as~\cite{Ajith:2011ec}:
\begin{equation}
\label{eq:TF2RShOfF}
h(f) \equiv \mathcal{C}\,f^{-7/6} \, \exp \left\{-\rmi \left[ \Psi(f) - \pi/4 \right] \right\},
\end{equation}
where $\mathcal{C}$ is a constant that depends on the relative sky-position and orientation of the binary with respect to the detector, and $f$ is the Fourier frequency. Note that we have kept only the leading term in the frequency-domain amplitude. The phase of the GW signal is given by
\begin{align}\label{eq:PsiofF}
\Psi(f) & = 2\pi f t_0 + \phi_0 + \frac{3} {128 \eta \, v_f^5} \left\{1 + v_f^2 \left[ \frac{55 \eta }{9}+\frac{3715}{756} \right] \right. \nonumber \\
    & \quad + v_f^3 \left[4\,\beta -16 \pi \right]   \nonumber\\
    & \quad + v_f^4 \left[\frac{3085 \eta ^2}{72}+\frac{27145 \eta }{504}+\frac{15293365}{508032} -10 \, \sigma_0 \right]  \nonumber\\
    & \quad + v_f^5 \left[\frac{38645 \pi}{756}-\frac{65 \pi  \eta }{9} -\gamma_0 \right] (3 \ln (v_f)+1)  \nonumber\\
    & \quad + v_f^6 \left[-\frac{6848 \gamma_E }{21}-\frac{127825 \eta ^3}{1296}+\frac{76055 \eta^2}{1728} \right.  \nonumber\\
    & \qquad \left. + \left(\frac{2255 \pi ^2}{12}-\frac{15737765635}{3048192}\right) \eta -\frac{640 \pi^2}{3} \right.  \nonumber\\
    & \qquad \left. +\frac{11583231236531}{4694215680}-\frac{6848 \ln (4v_f)}{21}\right]  \nonumber\\
    & \quad \left. + v_f^7 \left[-\frac{74045 \pi  \eta ^2}{756}+\frac{378515 \pi  \eta }{1512}+\frac{77096675 \pi }{254016} \right] \right\},
\end{align}
where $t_0$ is the time of arrival of the signal at the detector and $\phi_0$ the corresponding phase, $v_f$ is related to the Fourier frequency $f$ by $v_f \equiv (\pi m f)^{1/3}$, $m \equiv m_1 + m_2$ is the total mass and $\eta \equiv m_1 m_2/m^2$ is the symmetric mass ratio of the binary, and $\gamma_E$ is the Euler gamma.  The spin effects in the waveform are completely known up to 2.5PN order ($v^5$), and are described by the following parameters:
\begin{eqnarray}
\label{eq:SigmaGammaEpsilonCoeffs}
\beta  & = & 113 \, \chi/12 , \nonumber \\
\sigma_0 & = & \left(-\frac{12769 \, (4 \eta -81)}{16 \, (76 \eta -113)^2}\right)\, \chi^2 , \nonumber \\
\gamma_0 & = & \left(\frac{565 \left(17136 \eta ^2+135856 \eta -146597\right)}{2268 \, (76 \eta -113)}\right) \, \chi , \nonumber \\
\end{eqnarray}
where $\chi$ is called the \emph{reduced-spin parameter}, which is related to the individual spins of the binary by
\begin{equation}
\chi \equiv \chis + \delta \chia -\frac{76\eta }{113} \chis,
\end{equation}
where $\delta \equiv (m_1 - m_2)/m$ is the asymmetric mass ratio, and the symmetric- and antisymmetric combinations of the spins:
\begin{eqnarray}
\chis &\equiv& \frac{1}{2} \, \left(\frac{\bS_1}{m_1^2}  + \frac{\bS_2}{m_2^2} \right) \cdot \LNh, \nonumber \\
\chia &\equiv& \frac{1}{2} \, \left(\frac{\bS_1}{m_1^2}  - \frac{\bS_2}{m_2^2} \right) \cdot \LNh.
\label{eq:chisandChia}
\end{eqnarray}

\section{Computation of the template-space metric}
\label{sec:metric}
\subsection{Overview of the metric formalism}

This Section provides a brief overview of the metric formalism proposed by Owen~\cite{Owen:1995tm} for laying down templates in the parameter space.
The waveform template $h(f)$ defined in Eqs.~(\ref{eq:TF2RShOfF})--(\ref{eq:PsiofF}) is parametrized by a set of parameters $\bm \lambda \equiv \{\bm \lambda_\mathrm{intr}, \bm \lambda_\mathrm{extr} \}$ where $\bm \lambda_\mathrm{intr}$ are the \emph{intrinsic} parameters (such as $m, \eta$ and $\chi$) that are intrinsic to the binary, and $\bm \lambda_\mathrm{extr}$ are the \emph{extrinsic} parameters (such as $t_0$ and $\phi_0$). The \emph{match} between two neighboring templates in the parameter space is defined by
\begin{equation}
\label{eq:match}
\match(\blambda, \Delta \blambda) \equiv \mathrm{max}_{\Delta \blambda_\mathrm{extr}} \left< \hat{h}(f; \blambda), \, \hat{h}(f; \blambda + \Delta \blambda) \right>,
\end{equation}
where the angular brackets denote the inner product inversely weighted by the one-sided power spectral density $S_h(f)$ of the detector noise, called \emph{overlap}:
\begin{equation}
\left<a, b\right> = 2 \, \int_{\flow}^{\fcut} \frac{a(f)\,b^*(f) + a^*(f)\,b(f)}{S_h(f)}\,\df.
\end{equation}
Note that, in Eq.(\ref{eq:match}), the overlap is maximized only over the extrinsic parameters. The lower frequency cutoff $\flow$ is typically determined by the detector noise, which rises steeply below $\flow$ due to the seismic noise. The upper frequency cutoff $\fcut$ is due to the PN approximation breaking down when the binary approaches close separations (typically taken as the frequency of the innermost stable circular orbit).  Also, the ``hats'' denote normalized waveforms: $\hat{h}(f) \equiv h(f)/\sqrt{\left<h(f),h(f)\right>} $. Note that, since the computation of the match (Eq.~\ref{eq:match}) requires normalized templates, we can effectively set $\mathcal{C} = 1$ in Eq.~(\ref{eq:TF2RShOfF}), and hence the extrinsic parameters describing the location and orientation of the binary do not appear in the problem. 

We obtain a convenient approximate expression for the match between neighboring templates by Taylor-expanding the match about $\Delta \blambda = 0$. Since the match function has its maximum value of unity at $\Delta \blambda = 0$, there are no linear terms in the expansion, and truncating the expansion at second order, we get
\begin{equation}
\match(\blambda, \Delta \blambda) \simeq 1 -  g_{ij} \, \Delta \lambda^i \Delta \lambda^j
\label{eq:matchmetric}
\end{equation}
where
\begin{equation}
g_{ij} \equiv -\frac{1}{2} \, \left(\frac{\partial^2 \match}{\partial \Delta \lambda^i \, \partial \Delta \lambda^j}\right)_{\Delta \blambda = 0}
\end{equation}
can be interpreted as a metric in the parameter space. Thus the \emph{mismatch} between two neighboring templates has the interpretation of the proper distance in the parameter space~\cite{Owen:1995tm}:
\begin{equation}
\label{eq:mismatchFrmMetric}
1 - \match = g_{ij} \,  \Delta \lambda^i \Delta \lambda^j.
\end{equation}

A convenient way of computing the template-space metric $g_{ij}$ is by projecting the \emph{Fisher information matrix} $\Gamma_{ij}$ on to the subspace orthogonal to $\blambda_\mathrm{extr}$~\cite{Owen:1998dk}. The fisher information matrix (of normalized waveforms $\hat{h}(f)$) is defined as:
\begin{eqnarray}
\Gamma_{ab} & = &  \left< \partial_a\, \hat{h}(f; {\bm \lambda}), \partial_b \,\hat{h}(f; {\bm \lambda}) \right>
\label{eq:FisherMatrix}
\end{eqnarray}
where $\partial_a$ denotes a partial derivative with respect to the parameter $\lambda_a$. The indices $a$ and $b$ take values from 1 to 5 (including both intrinsic and extrinsic parameters).
The template space metric $g_{ij}$ can be computed by projecting $\Gamma_{ab}$ on to the subspace orthogonal to $\blambda_\mathrm{extr}$~\cite{Buonanno:2002ft}. i.e.,
\begin{equation}
\mathbf{g} = \bm \Gamma_1 - \bm \Gamma_2 ~  \bm \Gamma_3^{-1} ~ \bm \Gamma_4.
\label{eq:MetricFromFishMat}
\end{equation}
Above $\mathbf{g}$ is a matrix with elements $g_{ij}$, where $i$ and $j$ take values 1 to 3 (intrinsic parameters only). Similarly $\bm \Gamma_1, \bm \Gamma_2, \bm \Gamma_3, \bm \Gamma_4$ are the sub-matrices of the Fisher matrix:
\begin{eqnarray}
\bm \Gamma_1 \equiv \left[
\begin{array}{ccc}
\Gamma_{11} & \Gamma_{12} & \Gamma_{13} \\
\Gamma_{21} & \Gamma_{23} & \Gamma_{23} \\
\Gamma_{31} & \Gamma_{32} & \Gamma_{33}
\end{array}
\right],~~
\bm \Gamma_2 \equiv \left[
\begin{array}{cc}
\Gamma_{14} & \Gamma_{15}  \\
\Gamma_{24} & \Gamma_{25}  \\
\Gamma_{34} & \Gamma_{35}
\end{array}
\right], \\
\bm \Gamma_4 \equiv \left[
\begin{array}{ccc}
\Gamma_{41} & \Gamma_{42} & \Gamma_{43} \\
\Gamma_{51} & \Gamma_{53} & \Gamma_{53} \\
\end{array}
\right],~~
\bm \Gamma_3 \equiv \left[
\begin{array}{cc}
\Gamma_{44} & \Gamma_{45}  \\
\Gamma_{54} & \Gamma_{55}  \\
\end{array}
\right].
\end{eqnarray}

\subsection{Choice of coordinate system}

It is convenient to compute the metric in terms of a new set of variables $\{\thetaz, \thetat, \thetats\}$, which we call \emph{dimensionless chirp times}~\cite{Sathyaprakash:1994,Owen:1998dk}. The advantage is that, in this coordinate system, the metric components are slowly varying over the parameter space. The chirp mass ($\Mc \equiv  m \eta^{3/5}$), the symmetric mass ratio $\eta$ and the reduced-spin parameter $\chi$ can be written in terms of $\thetaz, \thetat$ and $ \thetats$ as
\begin{eqnarray}
\Mc  & = & \frac{1}{16 \, \pi f_0} \left(\frac{125}{2 \, \thetaz^3} \right)^{1/5}, \nonumber \\
\eta & = & \left(\frac{16 \, \pi^5}{25} \frac{\thetaz^2}{\thetat^5} \right)^{1/3}, \nonumber \\
\chi & = & \frac{48 \, \pi \, \thetats}{113 \, \thetat}.
\label{eq:MassSpinToThetas}
\end{eqnarray}
The dimensionless parameters $\thetaz, \thetat$ and $\thetats$ are related to the familiar chirp time~\cite{Sathyaprakash:1994} parameters $\tau_0, \tau_3$ and $\tau_\mathrm{3S}$ by
\begin{eqnarray}
\thetaz  = 2 \pi f_0 \tau_0, ~~ \thetat  = - 2 \pi f_0 \tau_3, ~~ \thetats  = 2 \pi f_0 \tau_\mathrm{3S},
\label{eq:thetatoTau}
\end{eqnarray}
where $\tau_0$ is the Newtonian chirp time, $\tau_3$ and $\tau_\mathrm{3S}$ are the spin-independent and spin-dependent terms of the 1.5PN chirp time, and $f_0$ is a reference frequency, such as the low-frequency cutoff of the detector sensitivity.

Using the relations given by Eq.~(\ref{eq:MassSpinToThetas}), the phase of the reduced-spin waveforms given in Eq.~(\ref{eq:PsiofF}) can be expressed in terms of $\thetaz, \thetat$ and $\thetats$ as
\begin{equation}
\Psi(f) = \sum_{k=0}^{k=8} \left[\psi_k + \psiL_k \, \ln\left(\frac{f}{f_0}\right) \right]\,\left(\frac{f}{f_0}\right)^{\frac{k-5}{3}},
\label{eq:PsiOfTTheta}
\end{equation}
where
\begin{eqnarray}
\psi_0 & = & \frac{3 \thetaz}{5}, \nonumber \\
\psi_1 & = & 0, \nonumber \\
\psi_2 & = & \frac{743 }{2016}   \left(\frac{25}{2 \, \pi^2 }\right)^{1/3}\, \thetaz^{1/3} \thetat^{2/3} + \frac{11 \pi \, \thetaz}{12 \, \thetat}, \nonumber \\
\psi_3 & = & -\frac{3}{2} (\thetat-\thetats), \nonumber \\
\psi_4 & = & \frac{675 \, \thetat \, \thetats^2 \, \left(8\times 10^{2/3} \pi ^{7/3} \thetaz^{2/3}-405 \, \sqrt[3]{10} \, \pi ^{2/3} \, \thetat^{5/3}\right)}{4 \sqrt[3]{\thetaz} \, \left(152 \, \sqrt[3]{10} \, \pi ^{5/3} \thetaz^{2/3}-565 \, \thetat^{5/3} \right)^2} \nonumber \\
	& + & \frac{15293365 \, \sqrt[3]{5} \, \thetat^{4/3}}{10838016 \times 2^{2/3} \pi ^{4/3} \sqrt[3]{\thetaz}} \nonumber \\
	& + & \frac{617 \, \pi ^2 \, \thetaz}{384 \, \thetat^2}+\frac{5429}{5376} \, \left(\frac{25 \pi \, \thetaz}{2 \, \thetat}\right)^{1/3},\nonumber \\
\psi_5 & = & \frac{140311625 \pi  \thetat^{2/3} \thetats}{180348 \left(565 \thetat^{5/3}-152 \sqrt[3]{10} \pi ^{5/3} \thetaz^{2/3}\right)} \nonumber \\
	& + & \frac{38645 \left(\frac{5}{\pi }\right)^{2/3} \thetat^{5/3}}{64512 \sqrt[3]{2} \thetaz^{2/3}}-\frac{732985\ 5^{2/3} \thetats \left(\frac{\thetat}{\thetaz}\right)^{2/3}}{455616 \sqrt[3]{2} \pi ^{2/3}} \nonumber \\
	& - & \frac{85 \pi  \thetats}{152 \thetat}-\frac{65 \pi }{384}+\phi_0, \nonumber \\
\psi_6 & = & \frac{15211\ 5^{2/3} \pi ^{4/3} \sqrt[3]{\thetaz}}{73728 \sqrt[3]{2} \thetat^{4/3}}-\frac{25565 \pi ^3 \thetaz}{27648 \thetat^3}-\frac{535 \gamma_E  \thetat^2}{112 \pi ^2 \thetaz} \nonumber \\
	& + & \left(\frac{11583231236531}{320458457088 \pi^2}-\frac{25}{8}\right) \frac{\thetat^2}{\thetaz}-\frac{535 \thetat^2}{336 \pi ^2 \thetaz} \, \ln \left(\frac{10 \thetat}{\pi \thetaz}\right) \nonumber  \\
	& + &    \left(\frac{2255 \sqrt[3]{5} \pi ^{5/3}}{1024\ 2^{2/3}}-\frac{15737765635 \sqrt[3]{\frac{5}{\pi }}}{260112384\ 2^{2/3}}\right) \sqrt[3]{\frac{\thetat}{\thetaz}}, \nonumber \\
\psi_7 & = & \frac{385483375 \sqrt[3]{5} \thetat^{7/3}}{173408256\ 2^{2/3} \pi ^{4/3} \thetaz^{4/3}}+\frac{378515\ 5^{2/3} \sqrt[3]{\frac{\pi}{2}} }{516096} \left(\frac{\thetat}{\thetaz}\right)^{2/3} \nonumber \\
	& - &  \frac{74045 \pi ^2}{129024 \thetat}, \nonumber \\
\psi_8 & = & 2 \pi f_0 t_0, \nonumber \\
\psiL_5 & = & \psi_5 - \phi_0, \nonumber \\
\psiL_6 & = & -\frac{535 \, \thetat^2}{336 \pi ^2 \, \thetaz}, \nonumber \\
\psiL_0 & = & \psiL_1 = \psiL_2 = \psiL_4 = \psiL_6 = \psiL_7 = \psiL_8 = 0.
\label{eq:PsiKTheta}
\end{eqnarray}
The advantage of using this coordinate system to describe the waveform is that, at least the lowest order terms have near linear dependence on these parameters. This means that the derivative of the waveform with respect to these parameters, and hence the template-space metric is slowly varying over the parameter space (see, e.g., Figure~\ref{fig:metric_determinants_chirptimes}). This makes the template placement problem considerably simpler.

\subsection{Template-space metric for reduced-spin templates}

\begin{figure*}[t]
\begin{center}
\includegraphics[width=6.5in]{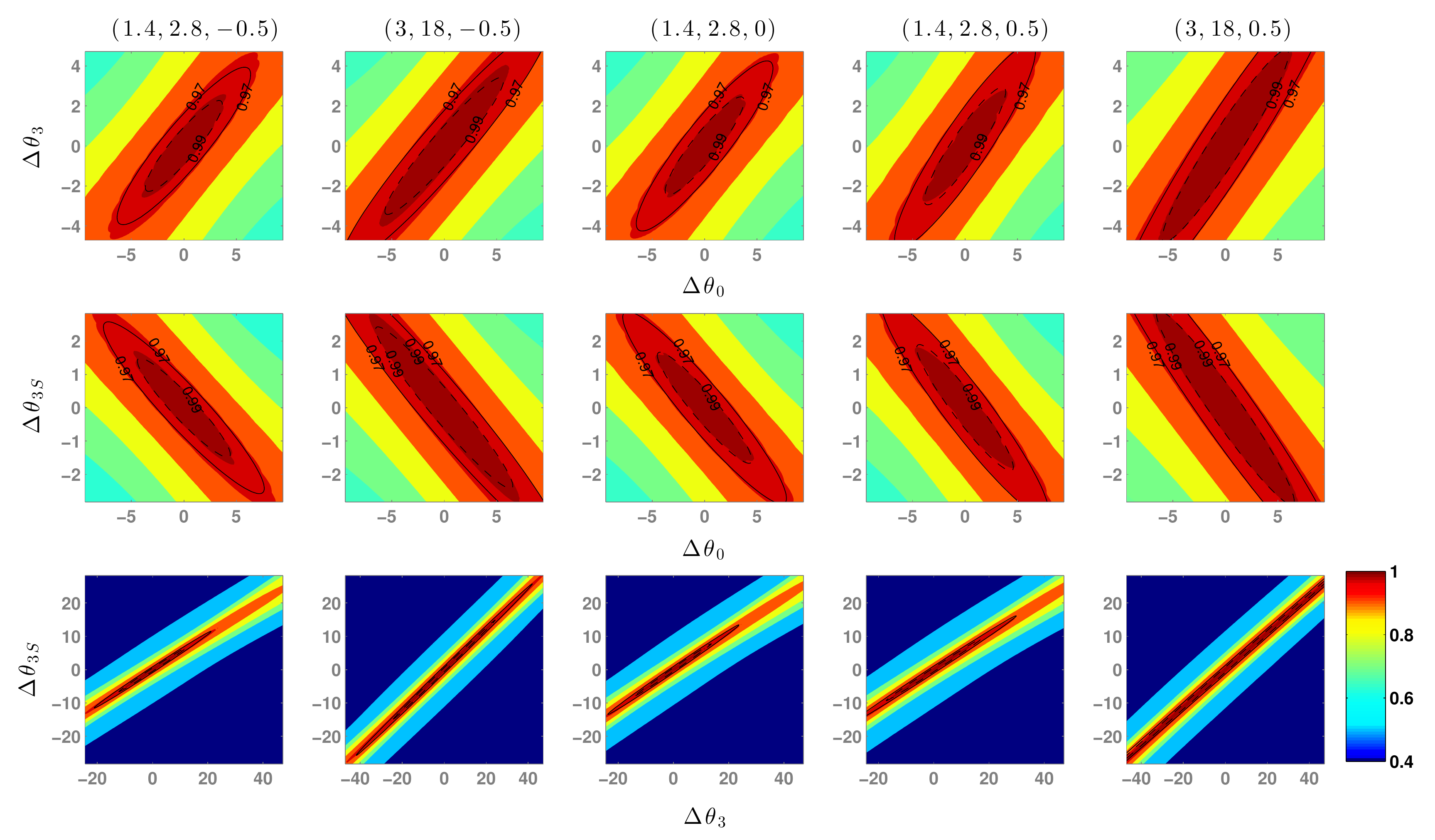}
\caption{Comparison of the match ellipses computed from the semi-analytical calculation of the metric (black ellipses) with contours of the match function  computed numerically (color contours). The component masses and reduced-spin parameter ($m_1,m_2,\chi$) corresponding to point in the parameter space relative to which the match function is computed is shown on the top of each column (masses in units of $M_\odot$). The different rows correspond to two-dimensional slices of these contours in the $\Delta \thetaz - \Delta \thetat$ plane (top row),  $\Delta \thetaz - \Delta \thetats$ plane (middle row) and $\Delta \thetat - \Delta \thetats$ plane (bottom row). The solid black ellipses correspond to a match of $0.97$ and dashed black ellipses correspond to a match of $0.99$.}
\label{fig:MatchContoursComparison}
\end{center}
\end{figure*}

Now we proceed to compute the template-space metric for the reduced-spin templates in the $\{\thetaz, \thetat, \thetats\}$ coordinate system. For the efficient computation of the metric, it is useful to define the following noise moments:
\begin{eqnarray}
\mathcal{I}_p \, (\fcut) & = & f_0^{-7/3} \int_{f_0}^{\fcut} \frac{df}{S_h(f)} \left(\frac{f}{f_0}\right)^{(p-17)/3}  \nonumber \\
\mathcal{J}_p \, (\fcut) & = & f_0^{-7/3} \int_{f_0}^{\fcut} \frac{df}{S_h(f)} \ln \left(\frac{f}{f_0}\right) \left(\frac{f}{f_0}\right)^{(p-17)/3} , \nonumber \\
\mathcal{K}_p \, (\fcut) & = & f_0^{-7/3} \int_{f_0}^{\fcut} \frac{df}{S_h(f)} \left[\ln \left(\frac{f}{f_0}\right)\right]^2  \left(\frac{f}{f_0}\right)^{(p-17)/3},
\end{eqnarray}
where $f_0$ is the low-frequency cutoff of the detector, $\fcut$ is the upper frequency cutoff of the template (typically the frequency of the innermost stable circular orbit, and hence depends on the masses) and $S_h(f)$ the one-sided power spectral density of the detector noise. The index $p$ can take values between 0 and 16, for a total of 17 values.

These noise moments have to be computed only once during the construction of a template bank. Then the metric at different points in the parameter space can be computed using these noise moments. Using Eqs.~(\ref{eq:FisherMatrix}), (\ref{eq:TF2RShOfF}), (\ref{eq:PsiOfTTheta}) and (\ref{eq:PsiKTheta}), the Fisher matrix can be computed as:
\begin{eqnarray}
\Gamma_{ab} & = & \frac{1}{2} \, \sum_{k=0}^{8} \sum_{\ell=0}^{8} \Biggl\{ \dera \psi_k ~ \derb \psi_\ell \, \mathcal{I}_{k+\ell} \, (\fcut)  \nonumber \\
    & + & \left[ \dera \psiL_k ~ \derb \psi_\ell + \dera \psi_k ~ \derb \psiL_\ell \right] \, \mathcal{J}_{k+\ell} \, (\fcut) \nonumber \\
    & + & \dera \psiL_k ~ \derb \psiL_\ell ~ \mathcal{K}_{k+\ell} \, (\fcut) \, \Biggr\}
\end{eqnarray}
where $\psi_k$ and $\psiL_k$ are given by Eq.~(\ref{eq:PsiKTheta}). From the Fisher matrix $\Gamma_{ab}$, the template space metric $g_{ij}$ can be computed using Eq.~(\ref{eq:MetricFromFishMat}). In this paper, we choose $\fcut$ as the frequency of innermost stable circular orbit of a test particle orbiting a Schwarzschild black hole: $\fcut = f_\mathrm{ISCO} = 6^{-3/2} (\pi m)^{-1}$.
\subsection{Comparison with numerical calculations}

\begin{figure*}
\centering
\includegraphics[width=7.1in]{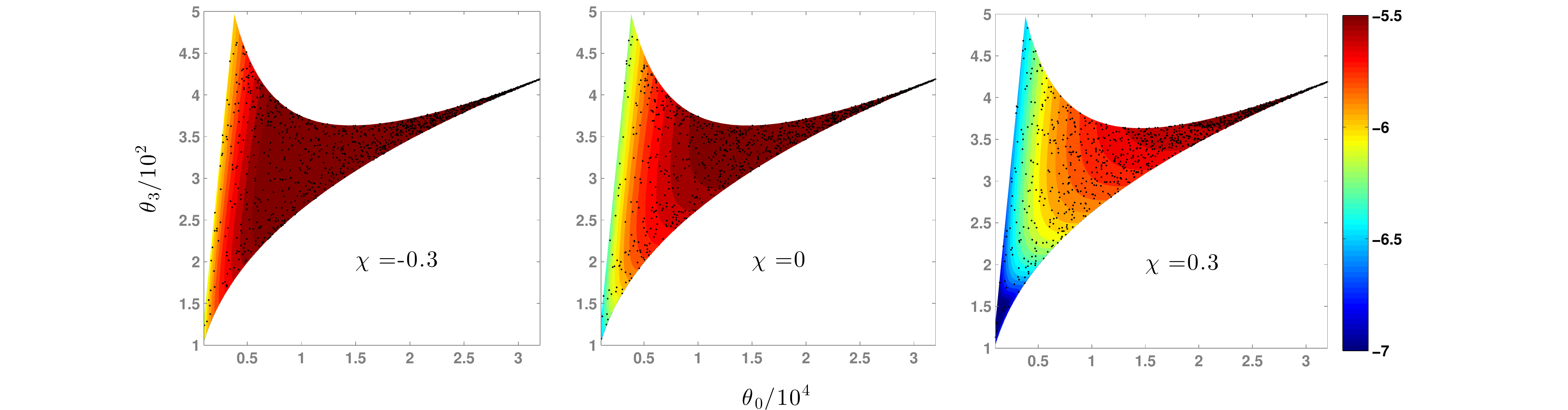}
\includegraphics[width=7.1in]{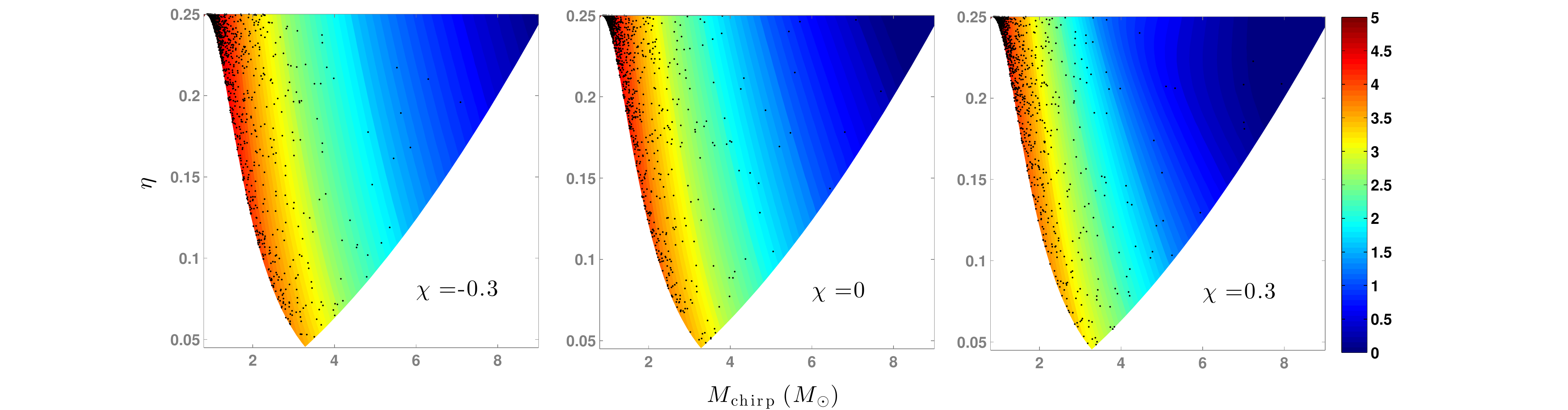}
\caption{The colored contours show the log of the square root of the determinant of the template space metric ($\log\sqrt{|\bg|}$) over the range of template bank parameters chosen in Table~\ref{tab:bank_params}. The number density of templates in each region of the bank is roughly determined by $\sqrt{|\bg|}$. The top panel corresponds to the metric computed in the dimensionless chirp time coordinate system $(\thetaz, \thetat, \thetats)$. The three subplots in the top row corresponds to three different slices in the 3-dimensional template bank, corresponding to $\chi = -0.3, 0, 0.3$. The black dots correspond to the templates placed in each slice (of thickness $\Delta \chi = 10^{-3}$) by the stochastic algorithm described in Section~\ref{sec:TemplateBank}. Notice that the template density in different regions generally agree with the expectation from the metric. In the $(\thetaz, \thetat, \thetats)$ coordinate system, the density has a maximum variation of $\sim 10$ over the entire template bank. In contrast, we show in the bottom panels the same quantity computed in the coordinate system described by $(\Mc, \eta, \chi)$. Notice that, in this coordinate system, the template density has considerable variation ($\sim 10^5$) over the parameter space. This illustrates the advantage of using the $(\thetaz, \thetat, \thetats)$ coordinate system.} 
\label{fig:metric_determinants_chirptimes}
\end{figure*}

Equation~(\ref{eq:matchmetric}) provides a very good approximation of the match $\match(\blambda, \Delta \blambda)$ between two neighboring templates $h(f; \blambda)$ and $h(f; \blambda + \Delta \blambda)$. But this approximation becomes inaccurate for the case of two templates placed at large distances in the parameter space (due to the inaccuracy of the truncated Taylor expansion). In this section, we compare the approximate match function computed using the metric with numerically exact computation of the match, as given by Eq.~(\ref{eq:match}). In the numerical computation of the match, maximization of the overlap over the extrinsic parameters $\blambda_\mathrm{extr} = \{t_0, \phi_0\}$ is performed using the standard techniques---maximization over $t_0$ is carried out by means of a Fast Fourier Transform, and the maximization over $\phi_0$ is carried out by employing two orthogonal templates~\cite{2012PhRvD..85l2006A}.

Figure~\ref{fig:MatchContoursComparison} shows the contours of the numerically computed match function (color-filled contours) along with the analytically computed match contours employing the metric (black ellipses). For each value of $(m_1, m_2, \chi)$, the corresponding  $\blambda \equiv (\thetaz, \thetat, \thetats)$ is computed by inverting Eq.~(\ref{eq:MassSpinToThetas}). The numerically computed match values of the template $h(f; \blambda)$ with neighboring templates $h(f; \blambda + \Delta \blambda)$ are reported by color-filled contours. Similarly, we compute the metric $g_{ij} (\blambda)$ at the point $\blambda$ corresponding to $(m_1, m_2, \chi)$. The contour in the  $\Delta \lambda \equiv (\Delta \thetaz, \Delta \thetat, \Delta \thetats)$ parameter space corresponding to a match value of 0.97 can be found by inverting Eq.~(\ref{eq:matchmetric}). This is an ellipsoid in this three dimensional parameter space. The black ellipses in Figure~\ref{fig:MatchContoursComparison} are the two-dimensional slices of the ellipsoids.

It can be seen from Figure~\ref{fig:MatchContoursComparison} that the analytically computed match ellipses based on the metric agree quite well with the exact, numerical computation of the match function. The minor disagreement between the two calculations is likely due to the fact that we are truncating the Taylor expansion of the match at quadratic order. But the approximate match ellipses are generally found to be smaller than the corresponding numerical exact match contours (this is consistent with the previous observations~\cite{Owen:1998dk}). This essentially means that the template bank constructed using the metric will slightly over-cover the parameter space.

Figure~\ref{fig:MatchContoursComparison} also demonstrates the advantage of
using the $(\thetaz, \thetat, \thetats)$ coordinate system for laying down
templates: Since the metric $\bg$ is significantly flatter over the parameter space,
the size and orientation of the match ellipses are very similar in all regions
in the parameter space. In Figure~\ref{fig:metric_determinants_chirptimes}, we
plot $\sqrt{|\bg|}$, which is proportional to the local ``density'' of templates
required to uniformly cover the parameter space. We see that $\sqrt{|\bg|}$
varies by a factor of only $\approx 10$ over $(\thetaz, \thetat, \thetats)$ space.
In contrast, a metric computed in the $(M_c, \eta, \chi)$ coordinate system 
has a variation of $\approx 10^5$ over the same parameter space (see Figure~\ref{fig:metric_determinants_chirptimes}). 
The near-uniformity of the chirp-time coordinates is a
desirable property for the speed of the template placement algorithm, which we 
discuss in the next section.

\section{Construction of the template bank}
\label{sec:TemplateBank}

\begin{table}
\centering
\begin{tabular}{c@{\quad}c}
\toprule
Bank parameter & Value \\
\midrule
Template waveform & \textsc{TaylorF2ReducedSpin} \\
Noise PSD model & \textsc{aLIGOZeroDetHighPower} \\
Low-frequency cutoff: $\flow$ & $20\,$Hz \\
Component mass: $m_1, m_2$ & $[1,\,20]\,M_\odot$ \\
Total mass: $m$ & $[2,\,21]\,M_\odot$ \\
NS spin: $\chi_i$ & $[-0.4,\,0.4]$ \\
BH spin $\chi_i$ & $[-0.98,\,0.98]$ \\
Minimum match: $\mathcal{M}_\text{min}$ & $0.95$ \\
Convergence criterion: $k_\text{max}$ & $1000$ \\
\bottomrule
\end{tabular}
\caption{Parameters used in generating the low-mass, reduced-spin template
bank.  The spin limits for black holes and neutron stars are different, corresponding to the different astrophysical expectations for the spins of these bodies. 
We consider a neutron star to be a body with mass $m_i \leq 2\,M_\odot$ and a black hole to be a body with mass $m_i > 2\,M_\odot$.}
\label{tab:bank_params}
\end{table}

\begin{figure}[t]
\begin{center}
\includegraphics[width=2.4in]{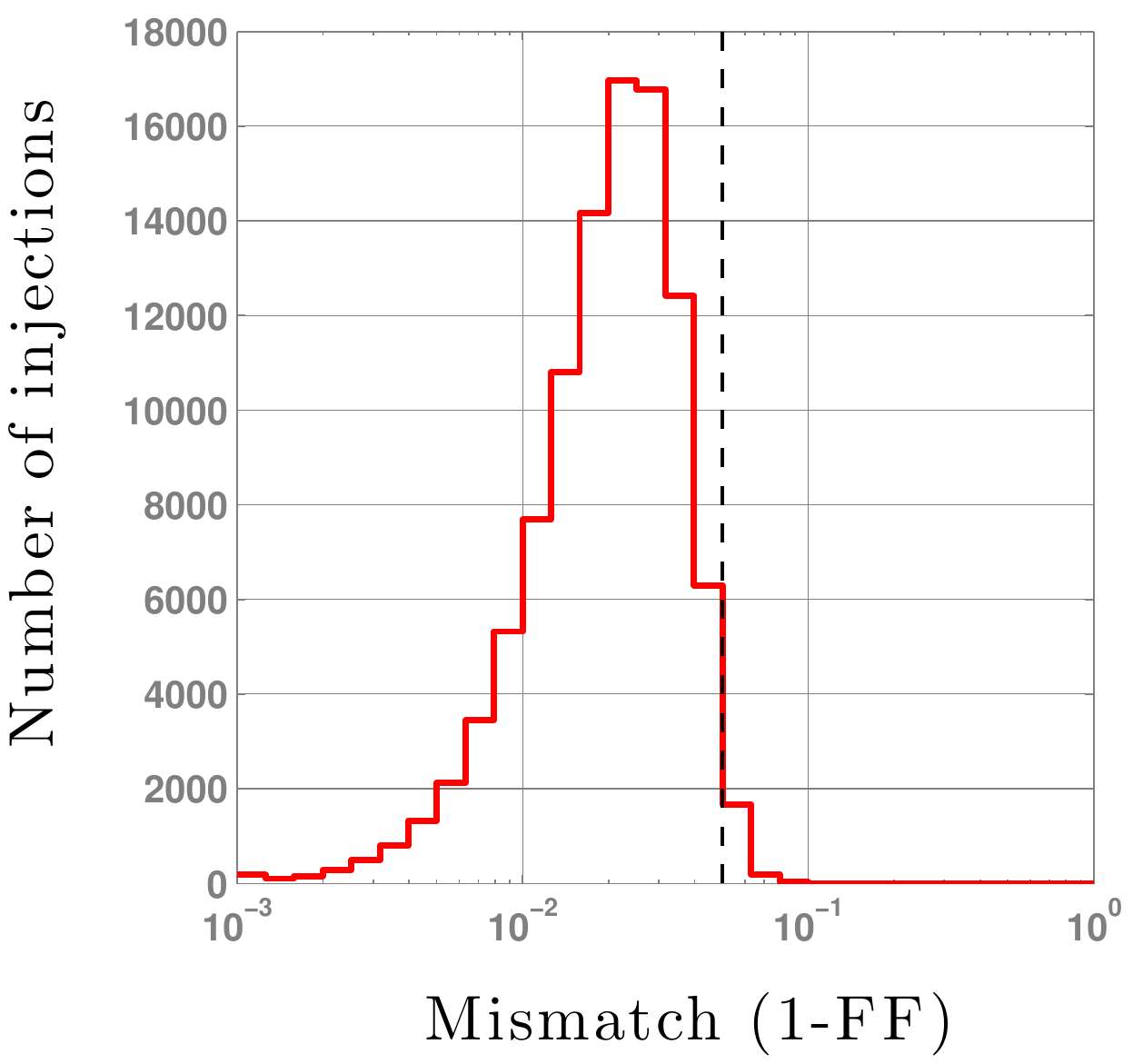}
\caption{Histogram of the achieved mismatch ($1 - \text{fitting factor}$) by the reduced-spin template bank in detecting reduced-spin injections . The plot is a demonstration of the achieved ``coverage'' of the bank. The vertical dashed line corresponds to a mismatch of 5\%. Only $\sim0.7\%$ of the $\sim 100,000$ injections have mismatch $> 5\%$. The effective fitting factor towards this population of injections is $\simeq 0.98$.}
\label{fig:BankCoverage}
\end{center}
\end{figure}

\begin{table}
\centering
\begin{tabular}{c@{\quad}c}
\toprule
Bank & Number of templates\\
\midrule
\SBank{} reduced spin & $549194$ \\	
\SBank{} non-spinning & $73275$ \\	
\bottomrule
\end{tabular}
\caption{Comparison of bank size between our reduced spin bank produced by \SBank{} with
the parameters of Table~\ref{tab:bank_params} compared to the non-spinning
\SBank{} version.} 
\label{tab:template_number}
\end{table}

We construct a template bank of the reduced spin waveforms using an implementation of the stochastic placement method proposed in~\cite{Harry:2009,2010PhRvD..81b4004M}. The method begins with a seed bank of template parameters $B_0 = \{\blambda^1_\mathrm{intr},\blambda^2_\mathrm{intr},\ldots,\blambda^N_\mathrm{intr}\}$, which may be empty~\footnote{Note that it is not necessary to provide discrete parameter values for the extrinsic parameters since the SNR of each template with the data is maximized over the extrinsic parameters using semi-analytical methods.}. A set of template waveform parameters $\blambda^\mathrm{prop}$ is proposed randomly and the bank $B_0$ is checked to see whether it already contains a template which sufficiently overlaps with the proposed template. We measure the coverage of the template bank via the \emph{fitting factor}~\cite{Apostolatos:1995pj}. The fitting factor is computed by maximizing the overlap of the proposed template $\blambda^\mathrm{prop}$ over the entire template bank $B_0$:
\begin{equation}
\mathrm{FF} = \max_{\blambda \,\in \,B_0} \match\,(\blambda,\blambda - \blambda^\mathrm{prop})\,,
\label{eq:FF}
\end{equation}
where $\match$ is the match function defined in Eq.~(\ref{eq:match}). The fitting factor gives the fraction of optimal SNR that can be obtained towards the proposed template waveform $\blambda^\mathrm{prop}$ using the existing template bank without including the newly proposed template. 

If the fitting factor for the proposed template is above a given \emph{minimum match} threshold $\mathcal{M}_\text{min}$, then the proposed template is discarded to prevent over-coverage and we repeat the process with the same bank seed $B_0$. Otherwise, the proposed template is added to the bank and we repeat the process using $B_1 \equiv B_0 \cup \{\blambda^\mathrm{prop}\}$ as the new bank seed. The process continues until some convergence criterion is satisfied. In our implementation, we terminate the bank construction when the mean number of discarded proposals per accepted proposal (averaged over the last ten accepted proposals) exceeds a specified critical value $k_\text{max}$.

The stochastic placement algorithm just described is straightforward to implement and is completely generic, independent of many of the particular details of the waveform family one is using. For example, while it is useful to have a metric that is nearly constant over the parameter space, this is not a strict requirement for the stochastic placement algorithm, as opposed to lattice-based approaches~\cite{Prix:2007ks}. We have implemented this algorithm with the generality of its use in mind. Thus, while in this paper we apply the stochastic bank placement code to the reduced-spin inspiral waveforms, the same code works with any waveform family, with or without an analytic approximation to the metric. We call the code \SBank{} and it is available for use in the \LALSuite{} data analysis package~\cite{LAL}. \SBank{} is implemented primarily in the Python programming language with speed-critical components in C\@.

For practical applications, we labored to make \SBank{} fast using some techniques suggested in related work~\cite{Harry:2009,2010PhRvD..81b4004M} and some novel.
\begin{itemize}
\item We obtain algorithmic speedup by testing each proposal against not the whole bank, but only its neighborhood of templates, defined by some fractional difference in $\theta_0$, the coordinate that is best fractionally measured; any template far away from the proposal in $\theta_0$ cannot have a high match.

\item In the intermediate to late stages of bank construction, we will sift through thousands of highly matching proposals before finding one with small enough match to accept it into our bank. Short-circuiting this search early saves enormous computation. Thus we stop computing further matches immediately upon finding a match greater than the target minimum match. So that we find the high matches even sooner, we evaluate matches in the order of increasing $\theta_0$ difference between the proposal and the bank seed.

\item Another technique we use is to draw proposals uniformly in $(\theta_0, \theta_3, \theta_\text{3S})$ space. As the true template density is proportional to $\sqrt{|\bg|}$ and $\bg$ is slowly varying in these coordinates, this reduces the number of proposals thrown at already over-tested regions of parameter space and puts them in under-tested regions.

\item Finally, the availability of the metric gives an analytic approximation to the mismatch, which significantly speeds up each iteration of this algorithm, but it is not strictly necessary.
\end{itemize}

Using our stochastic template placement code \SBank{}, we produced a template
bank with a reduced-spin dimension designed to capture astrophysically
plausible spins with the Advanced LIGO zero-detuned, high-power configuration%
~\cite{aLIGOPSD}. The template bank parameters are listed in
Table~\ref{tab:bank_params}. The waveform and noise model names refer to their
designations within the \textsc{LALSimulation} software library~\cite{LAL}.

We subjected this bank to a verification bank simulation, where we draw
simulation waveforms from the same reduced-spin waveform family discussed in Sec.~\ref{sec:Waveforms}
(called \textsc{TaylorF2ReducedSpin} in \textsc{LALSimulation}) with parameter ranges given in
Table~\ref{tab:bank_params} and we record the fitting factor of the bank towards each simulation
waveform. The simulation parameters are chosen uniformly in $(m_1, m_2, \chi)$ rather than
$(\theta_0, \theta_3, \theta_\text{3S})$. The results are shown in
Figure~\ref{fig:BankCoverage}. We see that in bulk, the algorithm has satisfied
the minimum match criterion with only a very small leakage.

Finally, in Table~\ref{tab:template_number} we compare our reduced-spin
template bank's size to non-spinning versions as generated by \SBank{}. 
Comparing the \SBank{}
reduced-spin bank to the \SBank{} non-spinning bank, we see that covering the
spin space of interest requires $\sim 7.5$ times the number of templates, and
approximately the same factor in matched filtering computational cost. In the
next section, we will see how much detection volume this extra computation
allows us to access.

\begin{table}[t]
\caption{Parameters used for the Monte-Carlo simulations of precessing PN binaries. A binary component is deemed a neutron star (NS) if its mass is  $\leq 2\,M_\odot$ and it is deemed a black hole (BH) if its mass is $> 2\,M_\odot$.}
\begin{center}
\begin{tabular}{c@{\quad}c}
\toprule
Simulation parameter & Value \\
\midrule
Waveform approximant & \textsc{SpinTaylorT5} \\
BH spin magnitudes: $||\bchi_i||$ & uniform($0,\,0.98$) \\
NS spin magnitudes: $||\bchi_i||$ & uniform($0,\,0.4$) \\
Cosine of spin orientations: $\LN^{\mathrm{ini}}.\bchi_i^{\mathrm{ini}}$ & uniform($-1,\,1$) \\
Cosine of sky location (polar):  $\cos \theta$ & uniform (-1, 1)\\
Sky location (azimuth): $\phi$ & uniform($0,\,2\pi)$ \\
Cosine of inclination angle: $\cos \iota$ & uniform($0,\,1$) \\
Polarization angle : $\psi$ & uniform($0,\,2\pi)$ \\
Luminosity distance: $d_L$ & 1 Mpc \\ 
Noise PSD model & \textsc{aLIGOZeroDetHighPower} \\
Low-frequency cutoff: $\flow$ & $20\,$Hz \\
\bottomrule
\end{tabular}
\end{center}
\label{tab:MonteCarloParams}
\end{table}%

\begin{figure*}[t]
\begin{center}
\includegraphics[width=3.5in]{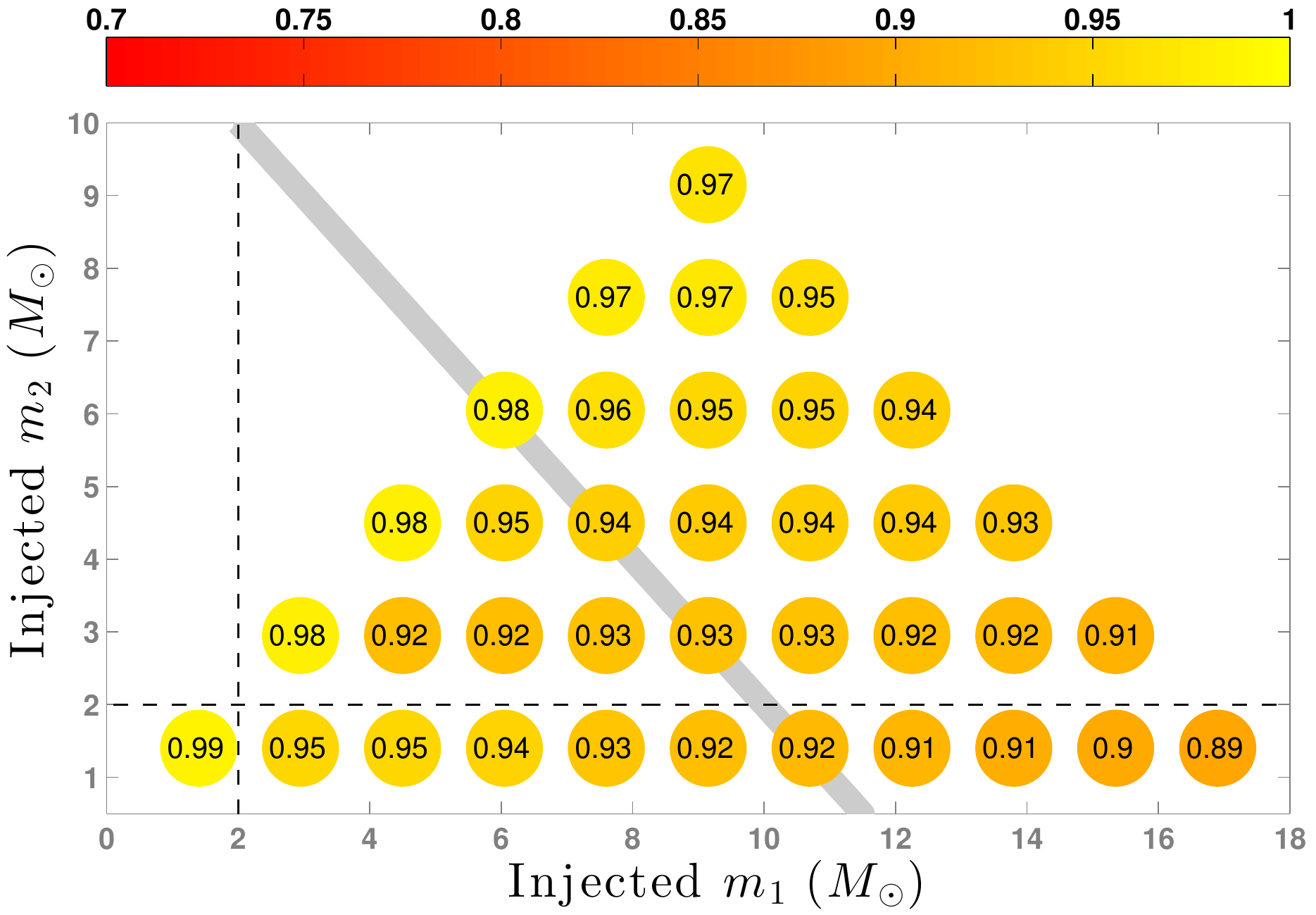}
\includegraphics[width=3.5in]{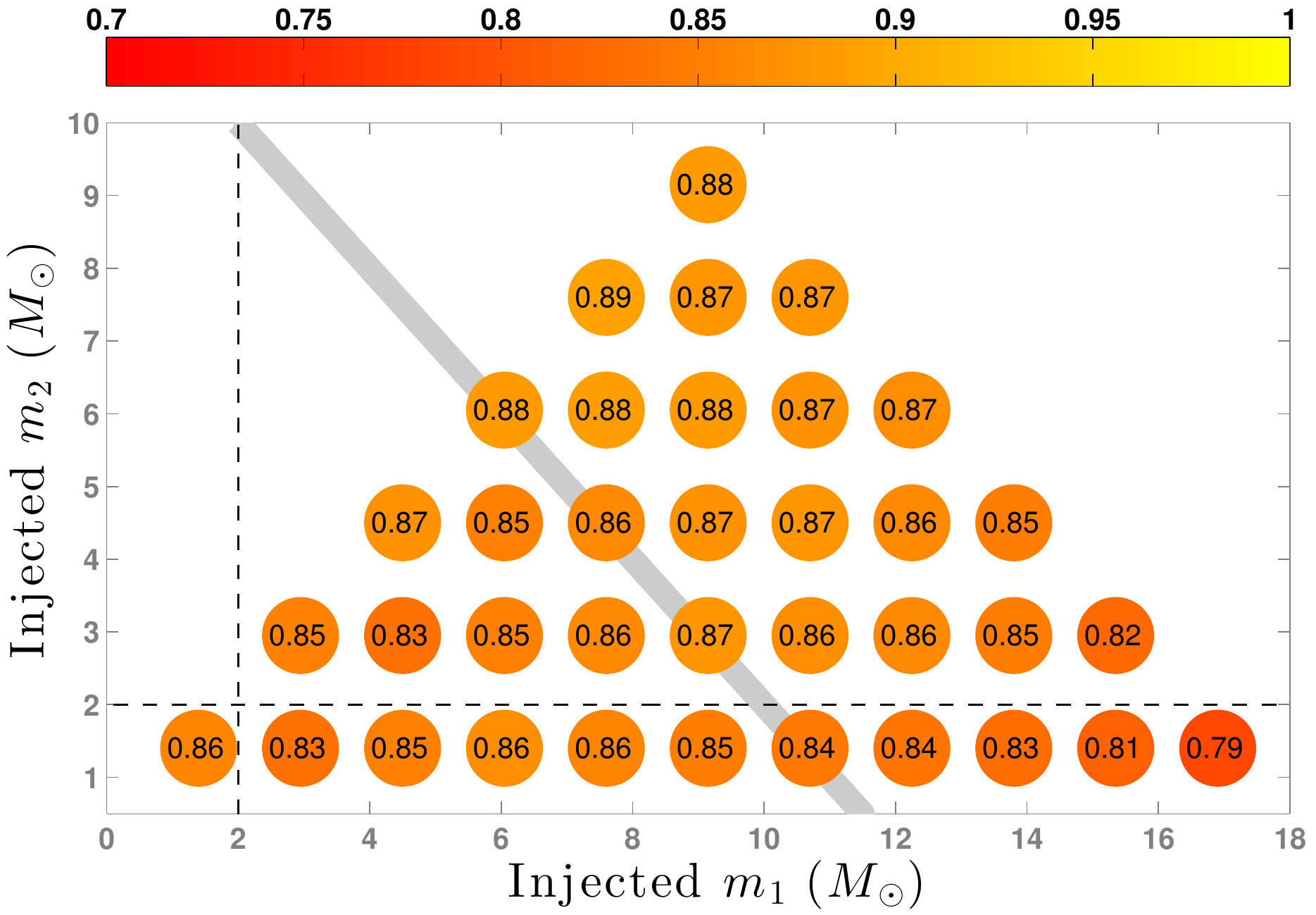}\\ \vskip -0.05in 
\advance\leftskip 0.16in \includegraphics[width=3.35in]{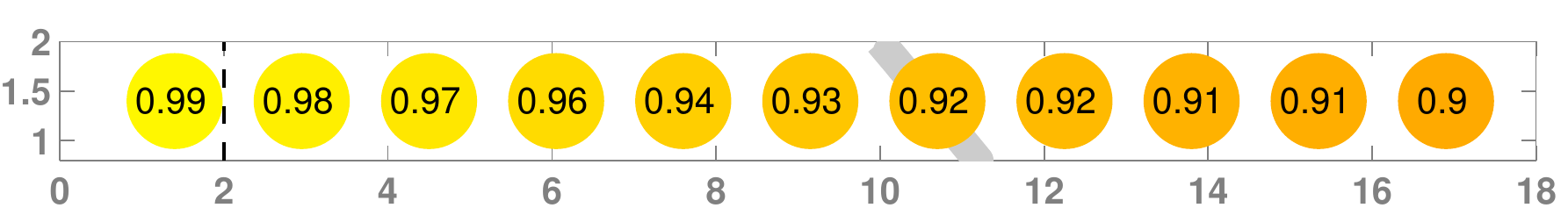}
\hskip 0.16in \includegraphics[width=3.35in]{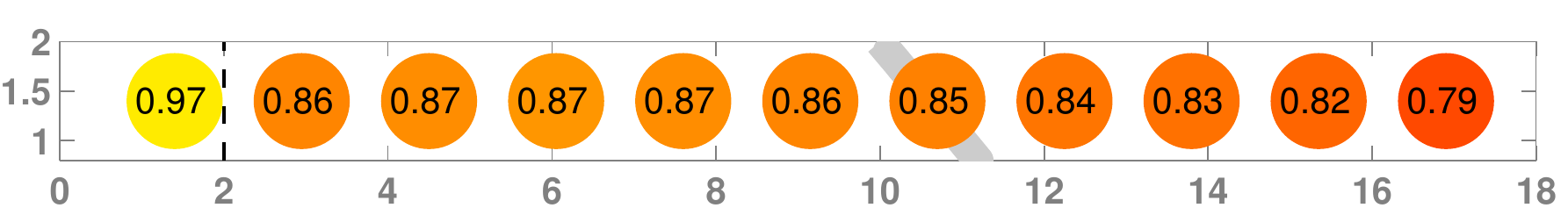}
\caption{The left plots show the effective fitting factor $\FFe$ of the reduced-spin template bank at different regions in the component mass plane. Each filled circle corresponds to 5000 injections (with fixed component masses, corresponding to the center of each circle) of generic spinning binaries with parameters reported in Table~\ref{tab:MonteCarloParams}. The gray line ($m_1 + m_2 = 12 M_\odot$) shows the expected boundary above which the contribution of merger-ringdown becomes non-negligible, and where the inspiral template bank needs to be replaced by an inspiral-merger-ringdown template bank. The black dashed lines correspond to the assumed boundary between the neutron-star- and black-hole mass. The top panel assumes that the maximum spin of neutron stars is 0.4, while the bottom panel assumes a maximum spin of 0.05 for neutron stars. The right plot shows the same for a non-spinning template bank.}
\label{fig:FFeSpinTaylorT5BankSim}
\end{center}
\end{figure*}

\begin{figure}[t]
\begin{center}
\includegraphics[width=3.5in]{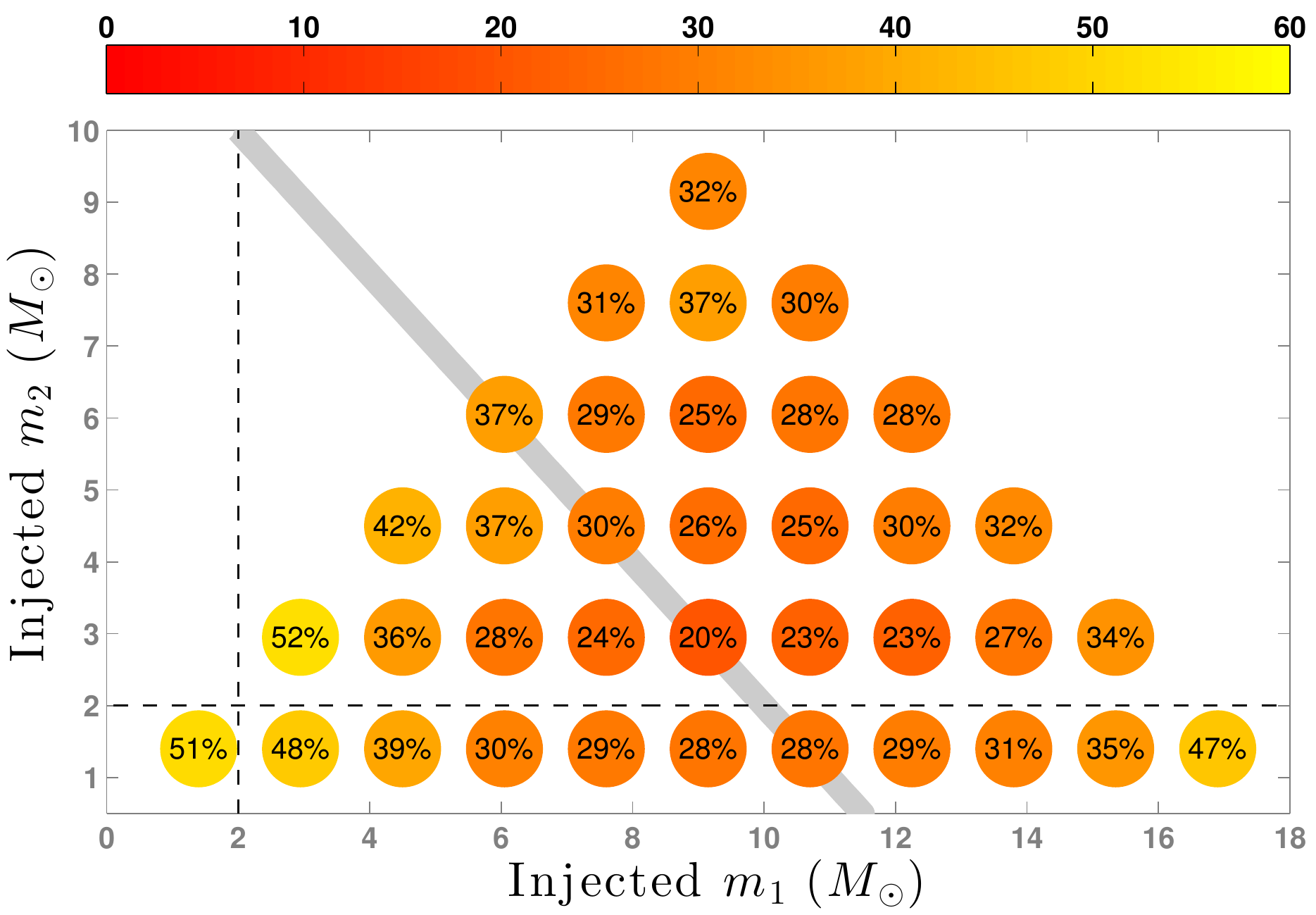}\\ \vskip -0.05in
\advance\leftskip 0.154in \includegraphics[width=3.33in]{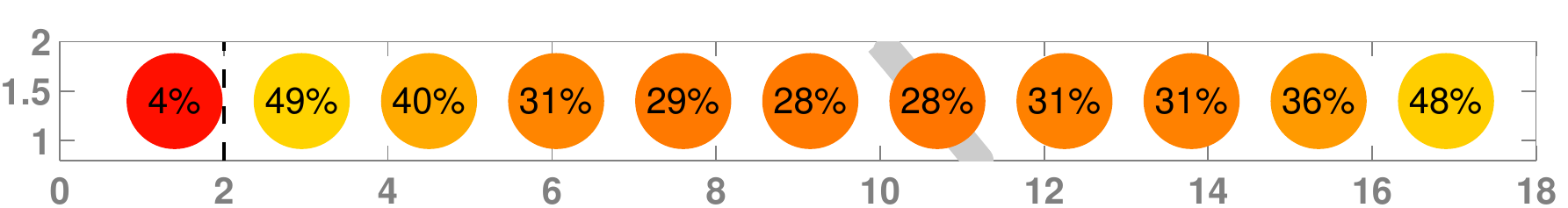}
\caption{Average increase in the detection volume of a search employing the reduced-spin template bank as compared to one employing a non-spinning template bank (corresponding to a fixed SNR threshold). The reduced-spin template bank is expected to bring about \red{$\sim20-52\%$} increase in the average detection volume (top panel), assuming that the maximum spin of neutron stars is 0.4. The bottom panel shows the result of the same calculation assuming a maximum spin of 0.05 for neutron stars.}
\label{fig:VolIncr}
\end{center}
\end{figure}

\section{Effectualness of the template bank in detecting generic spinning binaries}
\label{sec:Application}

In this section we study the \emph{effectualness}~\cite{DIS98} (a measure of the ability of a suboptimal template bank in detecting a family of target signals) of the reduced-spin template bank in detecting generic spinning binaries. We evaluate the effectualness by simulating a large number of inspiral signals from generic spinning binaries drawn from an astrophysical motivated distribution. For a given target waveform $h^{\mathrm{targ}}(f; \btheta)$, we compute the fitting factor for this signal against the bank using Eq.~(\ref{eq:FF}). The calculation has to be repeated over different values of $\btheta$, which describes the masses, spins and other parameters describing the relative location and orientation of the ``target binary'' with respect to the detector.

The intrinsic luminosity of the target binary as well as the fitting factor of the templates depend not only on the masses and spins, but also on the parameters describing the location and orientation of the target binary. For example, the modulational effects of precession are the highest for binaries highly inclined with respect to the detector, while the intrinsic luminosity of such binaries is lower (as compared to binaries which are nearly ``face on''). Thus, highly inclined binaries (which show the largest modulational effects of precession) are intrinsically less likely to be observed as compared to binaries that are face-on.

In order to take into account such selection effects in evaluating the effectualness of the template bank, we perform a Monte-Carlo simulation of generic spinning binaries and average the fitting factor over the population. The waveforms are generated by solving the ordinary differential equations given by Eq.~(\ref{eq:PNEvlEqns}) in the ``\textsc{TaylorT5}'' approximation (see Sec.~III of Ref.~\cite{Ajith:2011ec} for the full description)~\footnote{This particular approximant is chosen so as to disentangle the effect precession from the effect of the difference between different PN approximants; see Appendix~\ref{app:EffectualnessSTT4} for a discussion.}. The target binaries (for a set of fixed values of component masses) are uniformly distributed in volume throughout the local universe. Spin magnitudes are distributed uniformly between zero and a maximum value (see Table~\ref{tab:MonteCarloParams}) and the spin angles are isotropically distributed. Cosine of the angle $\iota$ describing the relative orientation of the initial total angular momentum of the binary with respect to the line of sight is uniformly distributed in the interval $(0, 1)$, while the polarization angle $\psi$ is uniformly distributed in $(0,\pi)$. A summary of the parameters of the Monte-Carlo simulations is given in Table~\ref{tab:MonteCarloParams}.

In order to evaluate the effectualness of the bank, we compute the \emph{effective fitting factor} $\FFe$~\cite{BCV2}, in the following way:
\begin{equation}
\FFe = \left( \frac{\overline{\rho^3_\mathrm{bank}} }{\overline{\rho^3} }\right)^{1/3},
\label{eq:FF_eff}
\end{equation}
where $\rho  \equiv \left<h^\mathrm{targ}, h^\mathrm{targ}\right>$ is the \emph{optimal} SNR in detecting the target binary, and $\rho_\mathrm{bank} \equiv \rho \, \FF $ is the \emph{suboptimal} SNR extracted by the template bank.  The bars indicate ensemble averages over the full parameter space (while keeping the component masses fixed). The effective fitting factor $\FFe$ describes average detection range by a suboptimal template bank as a fraction of the detection range using an optimal template bank. The corresponding fractional detection volume (and hence the fractional event rates assuming that the binaries are uniformly distributed throughout the universe) is given by the cube of $\FFe$.

The estimated effective fitting factor $\FFe$ of the reduced-spin template bank is shown in the left panel of Figure~\ref{fig:FFeSpinTaylorT5BankSim}. The figure suggests that the bank is effectual towards detecting generic spinning binaries over almost all the relevant regions in the ``low-mass'' parameter space ($m_1 + m_2 < 12\,M_\odot$)~\footnote{We conveniently define the ``low-mass'' range of the parameter space based on the previous studies using non-spinning inspiral waveforms, where it was shown that it is essential to include the effects of post-inspiral stages in the waveform for binaries with total mass $\gtrsim 12 M_\odot$~\cite{Buonanno:2009zt,Ajith:2007xh}.}. The effective fitting factor is always greater than \red{$\sim 0.92$}, and over a significant fraction of the ``low-mass''  parameter space the fitting factor is greater than $0.95$ (note that the minimum match requirement $\mathcal{M}_\mathrm{min}$ on the template bank was chosen to be 0.95). Note that the region above the gray line in the figure is the region where the contribution from the post-inspiral stages are expected to be significant, and the inspiral template bank needs to be replaced by an inspiral-merger-ringdown bank.

\begin{figure}[t]
\begin{center}
\includegraphics[width=3.5in]{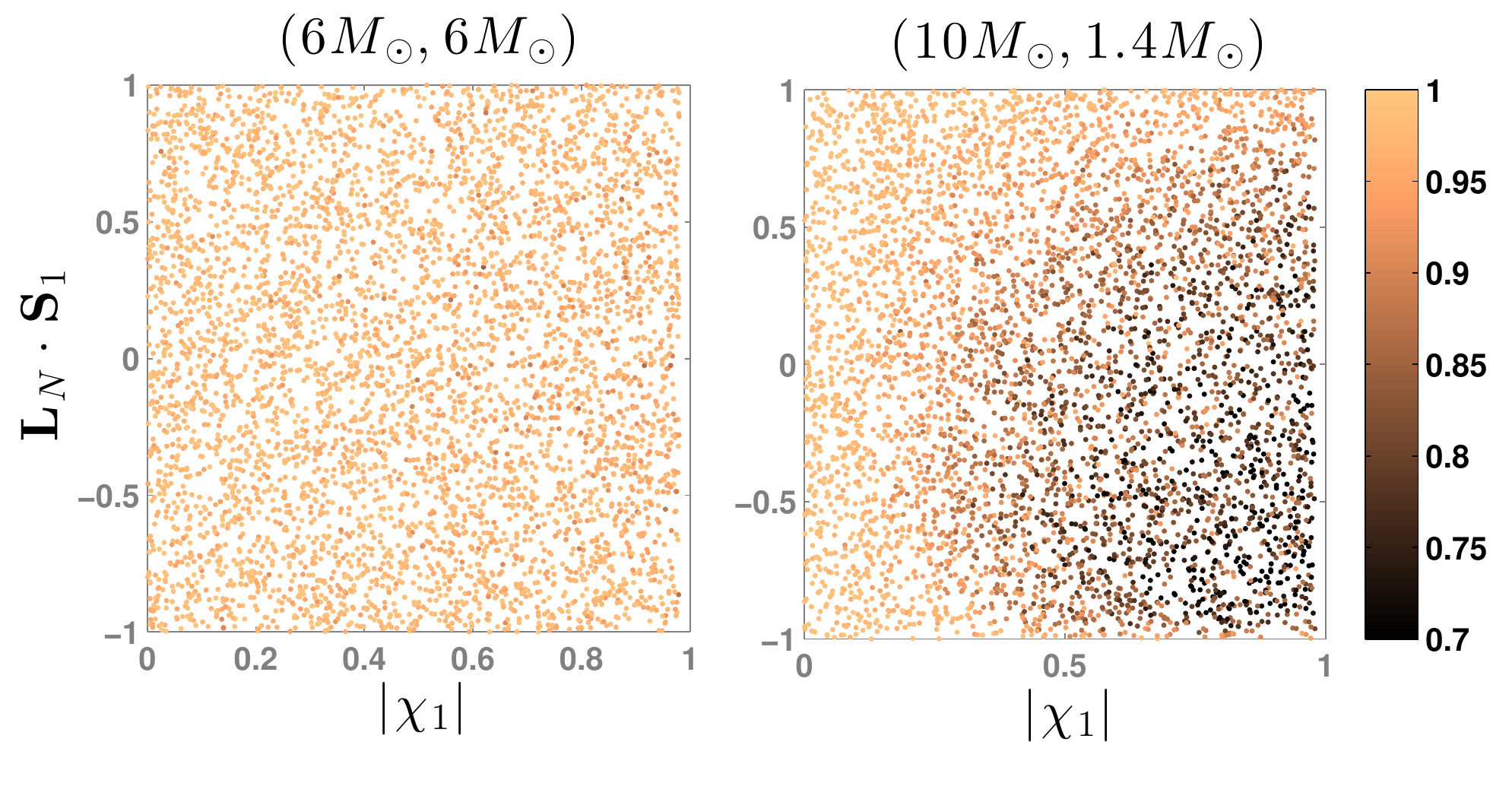}
\caption{Fitting factor (indicated by the color of the dots) of the reduced-spin template bank in detecting generic spinning binaries with component masses $(6 M_\odot, 6 M_\odot)$ [left plot] and $(10 M_\odot, 1.4 M_\odot)$ [right plot]. The x-axis corresponds to the spin magnitude of the more massive compact object, while the y-axis corresponds to the cosine of the angle between the spin and initial Newtonian orbital angular momentum. In the left plot (equal-mass binary) fitting factors are $\sim 1$ irrespective of the magnitude and orientation of the spin vector, while in the right plot (highly unequal-mass binary) fitting factors can be as low as $\sim 0.7$ for binaries with large, misaligned spins.}
\label{fig:FFScatterPlot_Chi1Mag_Chi1Angle}
\end{center}
\end{figure}

\begin{figure}[t]
\begin{center}
\includegraphics[width=3.5in]{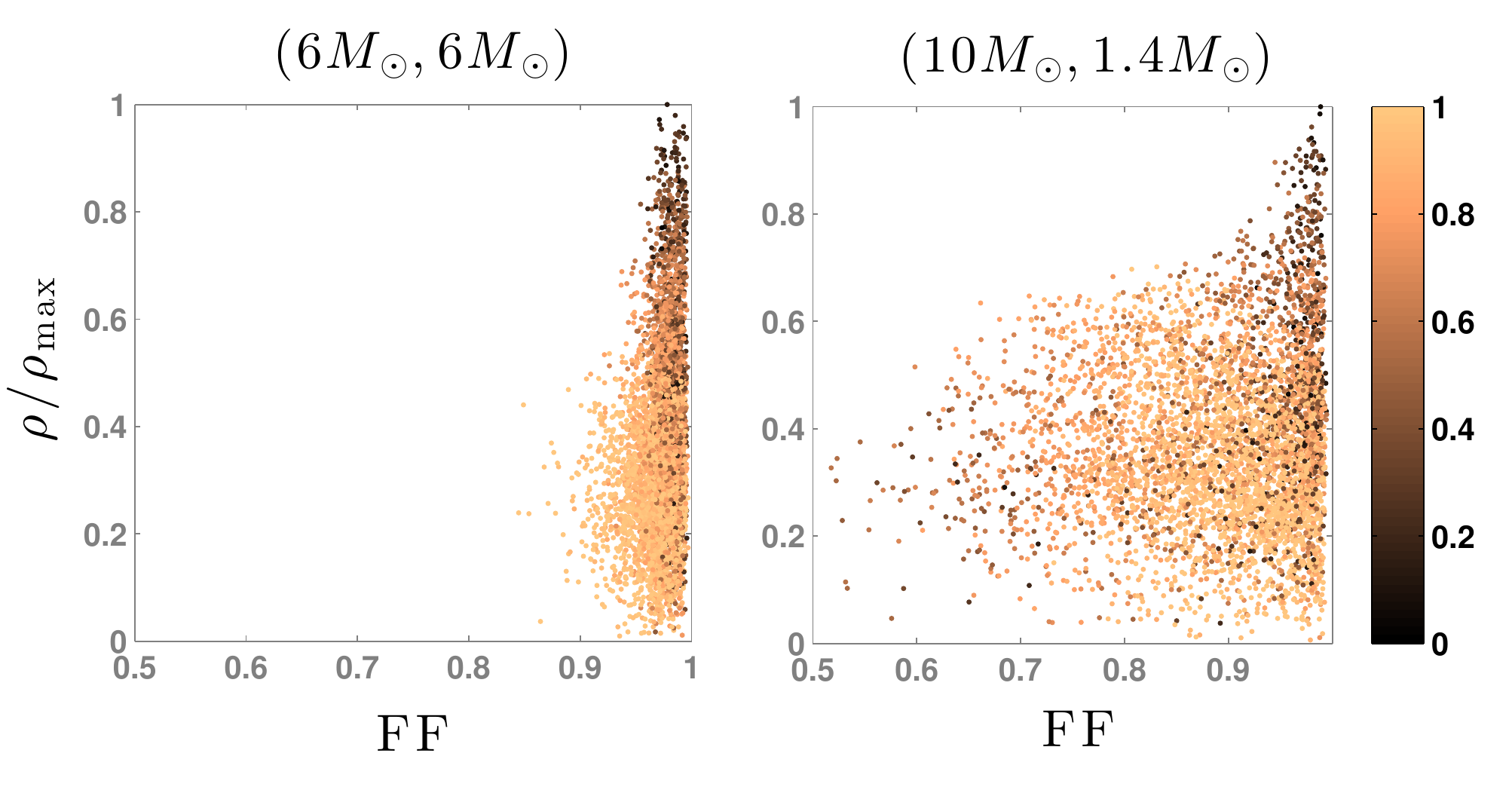}
\caption{Normalized SNR (such that the maximum SNR is 1) of generic spinning binaries plotted against the fitting factor (FF) of the reduced-spin template bank in detecting them. It can be seen that fitting factors are high towards binaries with large SNR. The color of the dots correspond to the sine of the inclination of the total angular momentum vector with respect to the line of sight (darker shades correspond to binaries whose total angular momentum is along the line of sight). The left plot corresponds to binaries with component masses $(6 M_\odot, 6 M_\odot)$ and the right plot to binaries with component masses $(10 M_\odot, 1.4 M_\odot)$.} 
\label{fig:ScatterPlot_SNR_FF_SinIncl}
\end{center}
\end{figure}

The high effectualness of the reduced-spin template bank towards generic spinning binaries can be attributed to two reasons. Firstly, for binaries with comparable masses ($m_1 \sim m_2$) the total angular momentum is dominated by the orbital angular momentum, and hence the modulational effects of spin precession on the orbit, and hence on the observed signal, is small. In this regime, non-precessing waveforms provide a good approximation to the observed signal. However, as the mass ratio increases, spin angular momentum becomes comparable to the orbital angular momentum and the modulational effects of precession become appreciable. Effectualness of non-precessing templates thus decrease with increasing mass ratio (see Fig.~\ref{fig:FFScatterPlot_Chi1Mag_Chi1Angle}). 

Secondly, there is an intrinsic selection bias towards binaries that are nearly ``face-on'' with the detector (where the modulational effects of precession are weak while the signal is strong) as opposed to binaries that are nearly ``edge-on'' (where the modulational effects are strong while the signal is weak). Thus the fitting factors are high towards binaries with large SNR. This effect is illustrated in Fig.~\ref{fig:ScatterPlot_SNR_FF_SinIncl} for the case of an equal-mass binary (left) and for the case of a highly unequal-mass binary (right). This helps the reduced-spin template bank to have reasonably high effective fitting factor towards a population of generic spinning binaries.

The reduction in the fitting factor of the reduced-spin template bank in the high-mass and high-mass-ratio regimes is due to multiple reasons. The modulational effects of precession increase with increasing mass ratio, which are not modeled by our templates. There are additional factors causing the loss: The difference between different PN approximants become considerable at the high-mass-, high-mass-ratio regime (reflecting the lack of knowledge of the higher order spin-dependent PN terms), causing appreciable mismatch between the target waveforms and the template waveforms even in regions where they should agree (e.g., in the limit of non-precessing spins). Hence, it is likely that the fitting factor can be further improved by including the higher order PN terms, assuming that these higher order terms will reduce the difference between different PN approximants (see, e.g.,~\cite{Nitz:2013mxa}). 

The effective fitting factor of a \emph{non-spinning} template bank (covering the same mass range) is shown in the right panel of Figure~\ref{fig:FFeSpinTaylorT5BankSim}. The fitting factor of the non-spinning bank is \red{$0.83$--$0.88$} over the same parameter space. The average increase in the detection volume provided by a search employing the reduced-spin template bank (as compared against the corresponding non-spinning template bank) is shown in Figure~\ref{fig:VolIncr}. The figure suggests that we can expect an increase of \red{$\sim 20$--$52\%$} in the average detection volume at a \emph{fixed SNR threshold}. Note that the real figure of merit of the improvement would be the increase in the detection volume for a \emph{fixed false-alarm rate}. Calculation of this requires the calculation of the increase in the false-alarm rate due to the increased number of templates in the bank. We leave this as future work.

Currently, all the \emph{observed} neutron stars in binaries have spin periods $\geq 22.7$\,ms~\cite{Kramer:2010hd}, which correspond to spin magnitudes of $||\bchi_i|| \lesssim 0.05$. While this is not necessarily an upper limit on neutron-star spins, this could be indicative of the typical spins. We have repeated the simulations by restricting the spin range of neutron stars in the target binaries to the interval (0, 0.05). This was found to make an appreciable difference only in the binary-neutron-star ($m_{1,2} \leq 2 M_\odot$) region of the parameter space. In this region, the effective fitting factor of the non-spinning template bank was increased to \red{$0.97$}. Thus, under this assumption, the non-spinning template bank appears to be adequate for the detection of GWs from binary neutron stars.

\section{Conclusion and outlook}
\label{sec:Conclusions}

Developing an effectual and computationally viable search for inspiralling binaries of spinning compact objects has been a long-standing problem in GW data analysis. The problem is made difficult by the large dimensionality of the parameter space. In this paper, we have attempted one of the first, albeit important, steps towards solving the problem: we have constructed a three-dimensional template bank that is effectual for the detection of a significant fraction of the generic spinning binaries in the ``low-mass'' parameter space. This development has been facilitated by a body of previous work: first, the realization that secular (non-precessing) spin effects are more important than the modulational effects for the case of comparable-mass binaries, which reduced the effective dimensionality of the problem into three~\cite{Ajith:2011ec}. The computation of closed-form templates modelling GWs from non-precessing-spin binaries that are parametrized in terms of a ``reduced-spin'' parameter has made it possible to compute the template-space metric in a semi-analytic fashion. Secondly, the demonstration of computationally efficient  stochastic-placement methods to place templates in the bank~\cite{Harry:2009,2010PhRvD..81b4004M} (which, as opposed to traditional lattice-based approaches, does not require the metric to be constant over the parameter space). 

We have demonstrated the expected effectualness of the template bank in the advanced detector era. For the spin distributions of the target binaries that we consider (see Table~\ref{tab:MonteCarloParams}), the effective fitting factor of the bank is in the range \red{$0.92$--$0.99$} over the ``low-mass'' binary ($m_1+m_2 \lesssim 12\,M_\odot$) parameter space. This is expected to bring about \red{$20$--$52\%$} increase in the detection volume as compared to a non-spinning template bank (for a fixed SNR threshold). The associated increase in the computational cost of the search would be roughly a factor of 7.5. Note that further optimization of the template-placement algorithm and the parameter ranges is possible to reduce the computational cost. This first demonstration of a template bank that is effectual (effective fitting factors \red{$> 0.92$}) over the entire parameter space of interest  promises a powerful and feasible method for searching for generic spinning low-mass binaries (including binary neutron stars, binary black holes and black-hole neutron-star binaries) in the advanced detector era. 

\appendix 
\section{Effectualness of the template banks against spinning waveforms generated in the TaylorT4 approximation}
\label{app:EffectualnessSTT4}

\begin{figure*}[t]
\begin{center}
\includegraphics[width=3.5in]{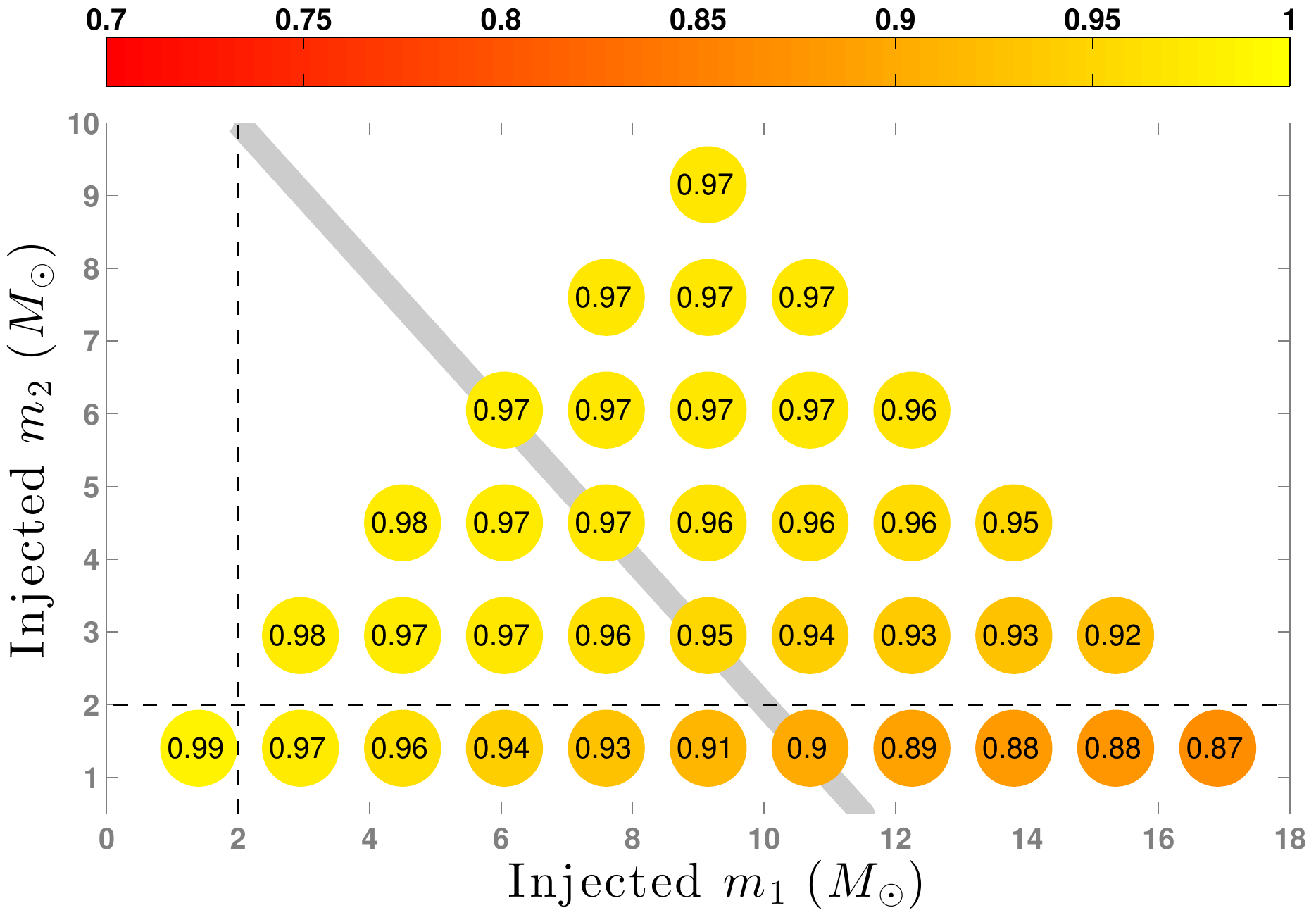}
\includegraphics[width=3.5in]{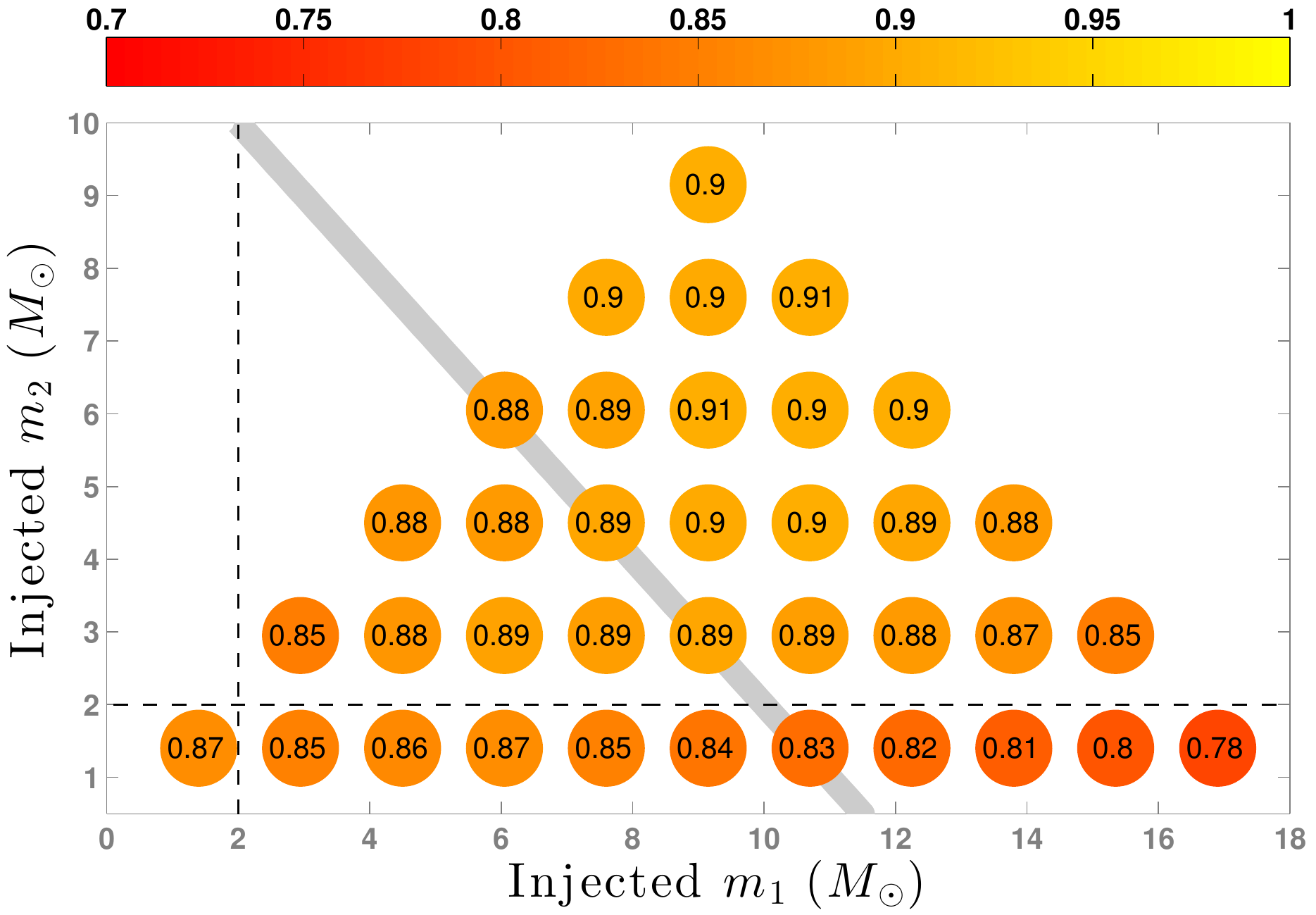}\\ \vskip -0.05in
\advance\leftskip 0.16in \includegraphics[width=3.35in]{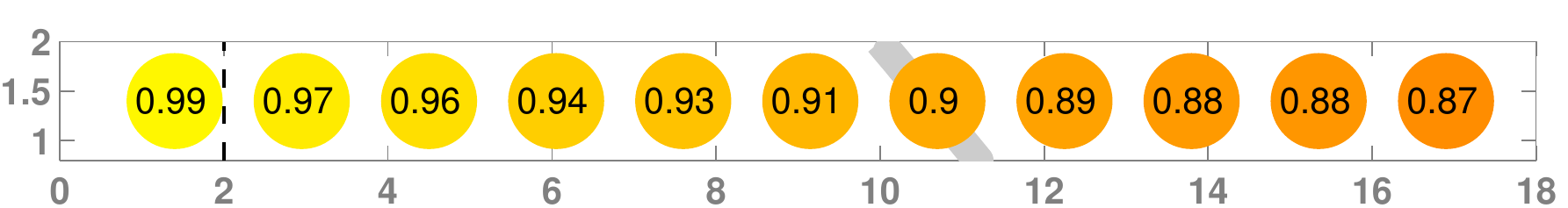}
\hskip 0.16in \includegraphics[width=3.35in]{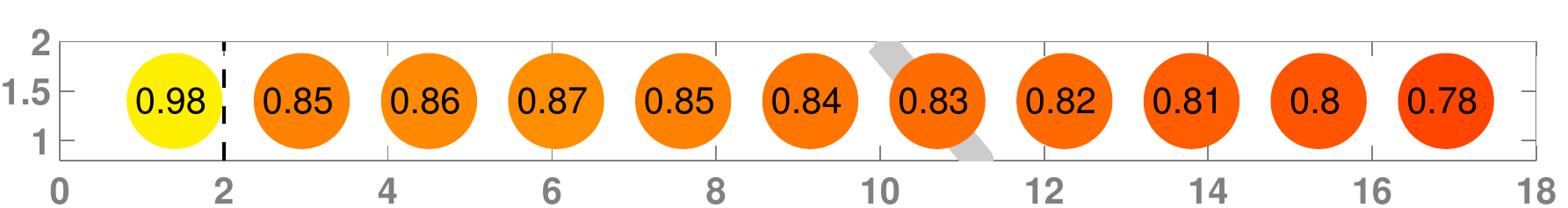}
\caption{Same as Fig.~\ref{fig:FFeSpinTaylorT5BankSim}, except that in this plot, the target waveforms are generated using the \text{TaylorT4} approximation. The difference in the effectualness between Fig.~\ref{fig:FFeSpinTaylorT5BankSim} and this figure is due to the difference between the two different PN approximants, and is a reflection of the current uncertainty in the PN waveforms. This could be improved by computing the higher order (spin-dependent) PN terms.}
\label{fig:FFeSpinTaylorT4BankSim}
\end{center}
\end{figure*}

\begin{figure}[t]
\begin{center}
\includegraphics[width=3.5in]{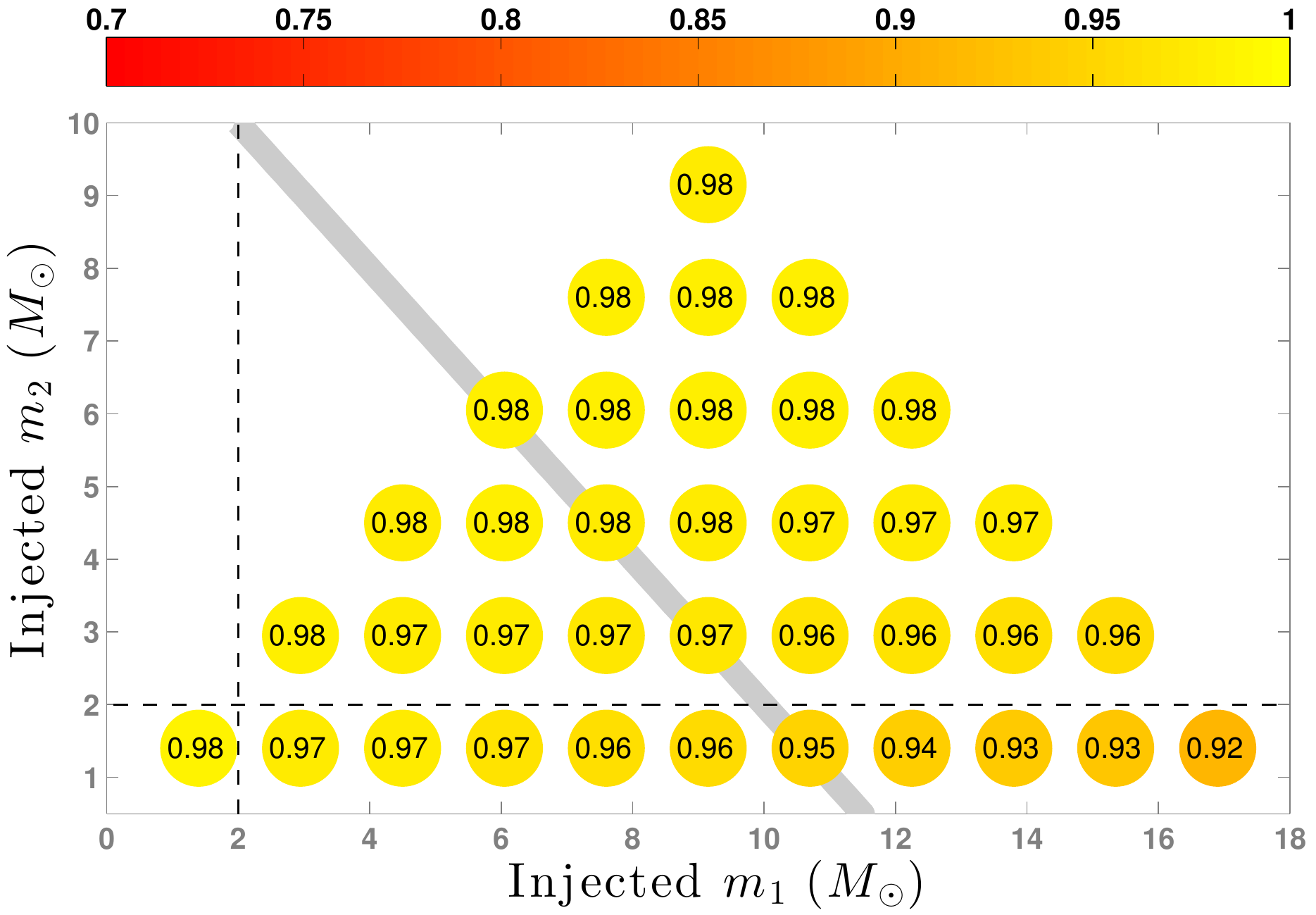} \\ \vskip -0.05in
\advance\leftskip 0.16in \includegraphics[width=3.35in]{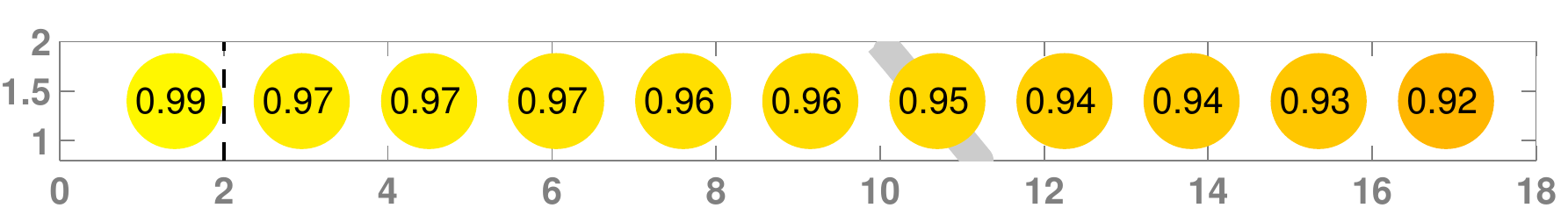}
\caption{Same as the left plot of Fig.~\ref{fig:FFeSpinTaylorT5BankSim}, except that in this plot, the target waveforms are \text{TaylorT4} with non-precessing spins.  It can be seen that the mismatch of the template bank at high mass ratios (in the low-mass regime) can be as large as 5\%, which cannot be attributed to the effects of precession.}
\label{fig:FFeSpinTaylorT4AlignedSpinBankSim}
\end{center}
\end{figure}

In Sec.~\ref{sec:Application}, we studied the effectualness of the reduced-spin template bank in detecting a population of generic spinning binaries assuming that the target signals are given by the \textsc{TaylorT5} PN approximant (see Sec.~III of~\cite{Ajith:2011ec} for the full description of this approximant). Note that this is one of the many approximations that can be used to compute PN waveforms from inspiralling compact binaries, and these different approximants can produce somewhat different results (see~\cite{Buonanno:2009zt} for an overview of different approximants). We would like to disentangle the loss of effectualness due to this effect from the loss due to the effect of precession. Thus, as the target waveform we need to use an approximant that is closest to the template in regions of parameter space where the target and template are expected to agree very well (e.g., in the limit of non-precessing spins). This is the motivation for choosing \textsc{TaylorT5} approximant as the target waveform. 

We do not expect \emph{a priori} one approximant to be closer to the signals given by nature than any other approximants. (This was further confirmed by comparisons of PN approximants with numerical-relativity simulations~\cite{MacDonald:2012mp, Hannam:2010ec}). Thus, in order to get a conservative estimate of the effectualness of the template banks, we compute their effective fitting factors towards signals from generic spinning binaries computed in the \textsc{TaylorT4} approximation (Fig.~\ref{fig:FFeSpinTaylorT4BankSim}). Note that the fitting factors at high mass ratios are slightly lower than what we see in Fig~\ref{fig:FFeSpinTaylorT5BankSim}. This difference arises from the fact that the waveforms computed using different approximants can be somewhat different, reflecting the current uncertainty in the PN waveforms (see, also,~\cite{Nitz:2013mxa} for a detailed discussion). It is likely that this uncertainty will decrease with the knowledge of higher PN terms (note that currently the spin-dependent terms are known only up to 2.5PN). 

We argue that one of the main reasons for the lower effectualness of the reduced-spin template bank towards \textsc{TaylorT4} waveforms at high mass ratios is, apart from the modulational effects of precession, the difference between PN approximants. In order to demonstrate this, we compute the effective fitting factor of the reduced-spin template bank towards \textsc{TaylorT4} waveforms \emph{with non-precessing} spins (Fig.~\ref{fig:FFeSpinTaylorT4AlignedSpinBankSim}). It can be seen that the mismatch of the template bank at high mass ratios (in the low-mass regime) can be as large as 5\%. This cannot be attributed to the effects of precession. These results greatly motivate the need of computing higher order spin terms in the PN approximation. 

\acknowledgments

The authors thank Chad Hanna for useful comments on the manuscript, and Duncan Brown, Alessandra Buonanno, Kipp Cannon, Gian Mario Manca, Chad Hanna, Ian Harry, Drew Keppel, Andrew Lundgren, Evan Ochsner, for useful discussions. This work is supported by the LIGO Laboratory, NSF grants PHY-0653653 and PHY-0601459, NSF career grant PHY-0956189 and the David and Barbara Groce Fund at Caltech. LIGO was constructed by the California Institute of Technology and Massachusetts Institute of Technology with funding from the National Science Foundation and operates under cooperative agreement PHY-0757058. PA thanks the hospitality of the Kavli Institute of Theoretical Physics (supported by the NSF grant PHY11-25915) during the preparation of this manuscript, while SP and NM thank the hospitality of the International Centre for Theoretical Sciences (ICTS). PA's research is also supported by a FastTrack fellowship and a Ramanujan Fellowship from the Department of Science and Technology, India and by the EADS Foundation through a chair position on ``Mathematics of Complex Systems'' at ICTS. NM is supported by the DST- MPG Max Planck Partner Group Grant (funded by the Department of Science and Technology, India and Max Planck Society, Germany) under the Grant no IGSTC/MPG/PG (!P) 2011. AN thanks the LIGO-REU program for support and Caltech for hospitality.  This paper has the LIGO Document Number LIGO-P1200106-v3.

\bibliography{SpinBank}

\end{document}